\documentclass[journal]{IEEEtran}

\usepackage{color}
\newcommand{\blue}[1]{\textcolor[rgb]{0,0,0}{#1}} 
\newcommand{\glue}[1]{\textcolor[rgb]{0,0,0}{#1}} 
 
\usepackage{amsfonts}
\usepackage{amssymb}
\usepackage{amsbsy} % Per a \boldsymbol
\usepackage{amsmath}
\usepackage{graphicx}
\usepackage{epstopdf}
\usepackage{amsthm}
\usepackage{bbm}

\newcommand{\tit}{Causal Inference in Geoscience and Remote Sensing from Observational Data}

\usepackage{array}
\newcolumntype{L}[1]{>{\raggedright\let\newline\\\arraybackslash\hspace{0pt}}m{#1}}
\newcolumntype{C}[1]{>{\centering\let\newline\\\arraybackslash\hspace{0pt}}m{#1}}
\newcolumntype{R}[1]{>{\raggedleft\let\newline\\\arraybackslash\hspace{0pt}}m{#1}}

\usepackage{pgffor} % for 
\usepackage{float} % avoid floats error latex
\usepackage{wrapfig}
\usepackage{flushend}  % balance last page columns

\usepackage{amsmath}
\usepackage{amssymb}
\usepackage{amsthm}
\usepackage{algorithm}
\usepackage{algorithmic}
\usepackage{placeins}
\usepackage{multirow}
\usepackage{color}
\usepackage{url}
\usepackage{cite}
\usepackage{wrapfig,lipsum,booktabs}
\usepackage[font=small,labelfont=bf]{caption}
\usepackage{graphicx}
\usepackage{xcolor}
\usepackage[all]{xy}
\usepackage{hyperref}
\usepackage{adjustbox}
\hypersetup{
    bookmarks=true,         % show bookmarks bar?
    unicode=false,          % non-Latin characters in Acrobat’s bookmarks
    pdftoolbar=true,        % show Acrobat’s toolbar?
    pdfmenubar=true,        % show Acrobat’s menu?
    pdffitwindow=false,     % window fit to page when opened
    pdfstartview={FitH},    % fits the width of the page to the window
    pdftitle={\tit},    % title
    pdfauthor={Adrian Perez-Suay and Camps-Valls},     % author
    pdfsubject={\tit},   % subject of the document
    pdfcreator={Adrian Perez-Suay and Camps-Valls},   % creator of the document
    pdfproducer={Adrian Perez-Suay and Camps-Valls}, % producer of the document
    pdfkeywords={Causality}{forward}{inverse}{modelling}{dependence}{causality}, % list of keywords
    pdfnewwindow=true,      % links in new window
    colorlinks=true,       % false: boxed links; true: colored links
    linkcolor={blue},          % color of internal links (change box color with linkbordercolor)
    citecolor=blue,        % color of links to bibliography
    filecolor=blue,      % color of file links
    urlcolor=blue           % color of external links
}
\usepackage{pgffor} % for 

\usepackage{setspace}
\newcommand{\Real}{\mathbb R}

\newcommand{\x}{{\mathbf x}}
\newcommand{\y}{{\mathbf y}}

\newcommand{\vect}[1]{{\boldsymbol{\mathbf{#1}}}} % vector
\newcommand{\mat}[1]{{\boldsymbol{\mathbf{#1}}}} % matrix

\newcommand{\s}{{\vect s}}

\newcommand{\dataset}{{\cal D}}

\newcommand{\param}{\mathbbm{z}}

\newcommand{\GP}[0]{\mathcal{GP}} 

\newcommand{\Normal}[0]{\mathcal{N}} 
 
\newcommand{\prob}{p}
 
\newcommand{\cut}[1]{} % cut out a part of the text

\title{\tit}
\author{Adri\'an P\'erez-Suay, Gustau Camps-Valls% <-this % stops a space
\thanks{Authors are with the Image Processing Laboratory, Universitat de Val\`{e}ncia, Spain (\url{http://isp.uv.es/}, e-mail: \url{[adrian.perez, gustau.camps]@uv.es}).}
\thanks{Research funded by the European Research Council (ERC) under the ERC-CoG-2014 SEDAL project (grant agreement 647423) and the Spainish Ministry of Economy, Industry and Competitiveness under the `Network of Excellence' program (grant code TEC2016-81900-REDT).}
}

\date{}

\begin{document}

\maketitle

\begin{abstract}
Establishing causal relations between random variables from observational data is perhaps the most important challenge in today's \blue{science}. In remote sensing and geosciences this is of special relevance to better understand the Earth's system and the complex interactions between the governing processes. 
In this paper, we focus on observational causal inference, thus we try to estimate the correct direction of causation using a finite set of empirical data. In addition, we focus on the more complex bivariate scenario that requires strong assumptions and no conditional independence tests can be used. %Tackling such problem in geoscience and remote sensing problems requires to adopt strong, and often unrealistic, assumptions. 
In particular, we explore the framework of (non-deterministic) additive noise models, which relies on the principle of independence between the cause and the generating mechanism. 
A practical algorithmic instantiation of such principle only requires 1) two regression models in the forward and backward directions, and 2) the estimation of {\em statistical independence} between the obtained residuals and the observations. The direction leading to more independent residuals is decided to be the cause. 
We instead propose a criterion that uses the {\em sensitivity} (derivative) of the dependence estimator, % to account for the asymmetry between the forward and inverse densities. 
the sensitivity criterion allows to identify samples most affecting the dependence measure, and hence the criterion is robust to spurious detections.
We illustrate performance in a collection of 28 geoscience causal inference problems, in a database of radiative transfer models simulations and machine learning emulators in vegetation parameter modeling involving 182 problems, %in the problem of identifying the atmospheric inversion layer from infrared sounding data, 
and in assessing the impact of different regression models in a carbon cycle problem. 
The criterion achieves state-of-the-art detection rates in all cases, it is generally robust to noise sources and distortions.
The presented approach confirms the validity in observational bi-variate problems in the Earth sciences.
\end{abstract}


\section{Introduction} \label{sec:intro}
%\begin{flushright}
%{\em Synoptikos, $\sigma\acute{\upsilon}\nu o \psi \iota\zeta$\\ 	
%``Affording a general view of a whole.''}
%\end{flushright}

\begin{flushright}
{\em ``... observational studies are an interesting and challenging field which demands a good deal of humility, since we can claim only to be groping toward the truth.''\\
\blue{William Cochran (1972)~\cite{Cochran65}.}}
\end{flushright}

% understanding is more difficult (and interesting) than predicting
\IEEEPARstart{T}{he} Earth is a highly complex and evolving networked system that we strive to understand better to deal with societal, economical and environmental challenges, such as climate change~\cite{IPCC14,Adam11}. There is an urgent need for tools that help us observe and study the Earth system. Nowadays, machine learning and signal processing play a crucial role for the production and analysis of Earth observation empirical data provided by a plethora of sensory systems and platforms. However, most statistical methods focus on {\em prediction} and estimation problems so they are only designed to take advantage of association relationships without considering causal mechanisms. %~\cite{biblia1,biblia2,biblia3,CampsValls16grsm}. 
Such methods provide little information about how variables actually interact with each other and, in this sense, are not very helpful to {\em understand} the underlying processes governing the system. The purpose of {\em causal inference} is precisely to go beyond association and to determine and discover links of causes and effects. Unlike association (e.g. correlation) studies, causal studies allow for understanding the underlying processes, and thus enable making inferences (i.e. predictions) of the effects of actions on the observed system~\cite{Pearl2000,SpirtesGlymourScheines2000,Peters18}. 

% Causal inference typically resort to models, but this is limited!
Causal inference should be ideally performed through the design of controlled experiments that try to avoid variable selection and confounding biases. Considering all possible variables and controlling all possible interactions would be of course the ideal scenario. Setting up such experiments is, however, not always possible, notably for ethical, economical or simply feasibility reasons~\cite{Rastetter1996}. This is often the case in empirical sciences, such as remote sensing, climate science and the geosciences, where one cannot control the whole set of variables affecting a given experiment. % over the Earth system. 
This is why actual causal experiments on the Earth system are often replaced by factorial experiments done with an ensemble of Earth-system model simulations~\cite{hegerl_use_2011}. Actually, a vast literature collectively perform {\em model-based} causal inference, and they focus on climate data only. The studies typically rely on climate models~\cite{attanasio_granger_2013}, and explore schemes for detection and attribution of plausible causes  by running models under different scenarios that consider or not the variable (forcing) under inspection. %Very often the objective is thus to determine the respective contribution of the different natural and anthropogenic forcings to the observed climate. %, with a huge focus on whether human activities has a significant impact on the climate. 

The main advantages of model-based approaches is that, in general, models encapsulate all the current knowledge about the system and thus they account for all the factors that can influence it. However, even though climate, biogeochemical or radiative transfer models are based on well-known physics equations, the combination of all internal processes and their couplings still make the interpretation of outputs very complicated, and it turns to be very difficult to disentangle internal from external induced variabilities. It is also worth noting that models are not a perfect representation of reality and many assumptions are made, so the eventual conclusions derived from these studies can be limited, or even wrong.
%The same holds true for local or regional-scale settings, such as studying relations at a leaf or canopy level for instance, where sophisticated biogeochemical and radiative transfer models are typically deployed~\cite{Reichstein10,Verrelst12rtm}.
%Nevertheless, all these experiments may be unreliable for causal inference depending on the model ability to represent the process of interest and are furthermore computationally very expensive. 

%%\pubidadjcol
\newpage

% causal inference from observations may help...
As an alternative to the use of models, one can resort to pure observational data, which opens a wide field in the current era of data deluge. It is acknowledged that the problem of inferring causation from empirical data has been traditionally considered unsolvable. 
Indeed, given two variables, identifying which is the cause and which one is the effect requires adopting (strong) assumptions. 
%For example, one should be aware of the possibly risk of not accounting for `confounders' (i.e. the existence of an unobserved variable that is the common cause of $X$ and $Y$), the risk of `selection bias' (i.e. the existence of common effects of both $X$ and $Y$), or combinations thereof. 
\blue{For example, it is assumed the absence of `confounding factors' that might drive both variables. This means that the system is fully described by two variables only, so there is not a third variable driving both. A second important assumption is known as the problem of `selection bias'. This implies that the observed variables should be representative of the causal relationship. A final third assumption commonly adopted is that no feedback loops can be found or created neither~\cite{Pearl2000}.}
A plethora of methods of causal discovery exist that try to remedy such limitations by 1) considering all potentially explanatory variables of the phenomenon, 2) selecting the most impactful ones, and 3) estimating {\em conditional independence} between (subsets of) observed variables to create directed graphs. Advances in observational causal inference have permitted to draw (partial) conclusions about the causal relationships in real-life problems~\cite{SpirtesGlymourScheines2000,Pearl2000,RichardsonSpirtes2002,Zhang2008}. 
{In remote sensing and geosciences, this is of special relevance to better understand the Earth's system and the complex and elusive interactions between the involved processes. Answering key questions may have deep societal, economical and environmental implications~\cite{Walther02,Adam11}.}

%In the last decades, however, many advances in the fields of statistics, machine learning, and signal processing allowed inferring causation with high and significant accuracy under some assumptions~\cite{SpirtesGlymourScheines2000,Pearl2000,RichardsonSpirtes2002,Zhang2008,Mooij16jmlr,Peters18}. 
%Establishing causal relations between random variables from observational data is perhaps the most important challenge in today's Science. 
%Recent developments, however, have shown high and significant accuracy in detection over representative datasets with ground truth~\cite{SpirtesGlymourScheines2000,Pearl2000,RichardsonSpirtes2002,Zhang2008,Mooij16jmlr}.

In the field of remote sensing and geosciences, two main methodological approaches exist, and \blue{both} of them consider the {\em multivariate} setting and that {\em time} is involved. Time gives an obvious intuition on causality ({\em ``the cause should precede the effect''}) and having access to multiple time series allows to overcome the selection bias problem. The seminal work by Granger~\cite{Granger69} proposed a rather simple statistical concept of causality based on prediction, and has been applied in many fields of Earth system science: 
%Granger, C. W. J. 1969 Investigating causal relations by econometric models and cross-spectral methods. Econometrica 37, 424-438. 
%The concept of {\em Granger's causality} has been exploited 
to perform attribution of climate change~\cite{Triacca2005, smirnov_granger_2009, leggett_granger_2015,attanasio_granger_2013}; 
to study the feedback mechanism between soil moisture and precipitation~\cite{salvucci_investigating_2002};
or to identify the relationship between sea surface temperature (SST) and the North Atlantic Oscillation (NAO)~\cite{mosedale_granger_2006}. Recently, Granger causality has been adapted to consider {\em nonlinear} relations in vegetation dynamics~\cite{Papagiannopoulou17}. 
%Papagiannopoulou C, Miralles D G, Verhoest N E C, Dorigo W A and Waegeman W 2017 A non-linear Granger causality framework to investigate climate vegetation dynamics. Geosci. Model Dev. 10 1945–60
As an alternative to prediction, a second family of methods consider that it is sometimes possible to retrieve, at least partially, the causal structure %encoded in the direct acyclic graph (DAG) of a graphical causal model 
using conditional independence tests between the variables in {\em PC schemes}~\cite{pc_1993}. The family is called {\em constraint-based search}, and has been used to study the causal interactions between climate modes of variability~\cite{ebert-uphoff_causal_2012}, as well as to construct climate causal networks~\cite{runge_identifying_2015,ebert-uphoff_new_2012}. Similar methodologies were applied to study the causal relationships with the Atlantic meridional overturning circulation (AMOC) \cite{schleussner_role_2014}, to analyze potential drivers of Artic Oscillation \cite{kretschmer_using_2016}, or to investigate the interactions between El Ni\~no-Southern Oscillation (ENSO) and the Walker circulation \cite{runge_quantifying_2014}. 

% we focus only on two variables! and no time is involved!
In this paper, unlike in the previous approaches, we will focus on the more challenging problem of inferring causality from observational data with two important constraints:
\begin{itemize}
    \item {\em Bivarate case.} We will consider the case of having access to two variables only, hence no conditional independence tests can be computed to guide the causal identification. % \red{Therefore, Reichenbach's common cause principle is not verifiable.}
    \item {\em Time is not explicitly considered.} We will not consider time-series explicitly in general, hence we deal with the problem of {\em `instantaneous causality'}. 
    \blue{This hampers the adoption of the `causes precede effects' rationale, so %here %the natural intuition of     Hume's reasoning cannot be adopted, nor 
    specific time-series approache such as those based on Granger causality methods cannot be applied.} Note that, however, time is often not necessary to discuss concepts such as statistical dependence and, in causal models, time is often not even needed to discuss the effect of interventions.
\end{itemize}
We call this setting the {\em bivariate instantaneous} case, which has been recently treated in~\cite{Mooij16jmlr,Peters18}. Let us thus assume that only two variables, $x$ and $y$, have been observed, and we aim to distinguish $x$ causing $y$ (indicated as $x\to y$) from $y$ causing $x$ (that is $y\to x$) using only purely observational data, i.e., a finite i.i.d.\ sample drawn from the joint distribution $p(x,y)$. Following the Bayes' theorem, there are two admissible partitions of the joint distribution:
$$p(x,y) = p(y|x)p(x) = p(x|y)p(y),$$
and the question is to decide which one is the causal one. The first one describes variable $x$ and the conditional $p(y|x)$ that can be interpreted as a function that translates the information contained in $x$ to $y$, so it assumes that $x$ causes $y$. The second decomposition assumes that modeling $y$ and the conditional $p(x|y)$ better explains the joint distribution, thus assuming that $y$ causes $x$. A possible answer to this question could be obtained through an intervention analysis (alter a variable and see the plausibility of the effect on the system), but this is not possible in our Earth science problems for obvious reasons. An alternative pathway is to rely on the {\em principle of independent mechanisms}~\cite{Spirtes,Peters18}: if we assume $x\to y$ so the causal partition is $p(x,y)=p(y|x)p(x)$, one would expect the conditional density $p(y|x)$ to provide no information about the marginal density $p(x)$, or at least less information than $p(x|y)$ would provide about $p(y)$. Therefore, a solution to the bivariate case reduces to estimate independence between the cause density and the mechanism producing the effect distribution. This is called the {\em `independence of cause and mechanism'} (ICM)~\cite{Daniusis10,Shajarisales15}, which is the one adopted in this paper. We should note that not all systems satisfy this principle. %, e.g. those tuned to each other by design or evolution. 
The problem is to achieve good density estimates of the conditionals and potential causes, and to measure independence between them accurately. 

%\red{A variety of causal discovery methods has been proposed in recent years that were claimed to be able to solve this task under certain assumptions\cite{FriedmanNachman2000,KanoShimizu2003,ShimizuHoyerHyvarinenKerminen2006,SunJanzingSchoelkopf2006,SunJanzingSchoelkopf2008,HoyerJanzingMooijPetersSchoelkopf_NIPS_08,MooijJanzingPetersSchoelkopf_ICML_09,ZhangHyvarinen2009,JanzingHoyerSchoelkopf2010,Mooij_et_al_NIPS_10,Daniusis_et_al_UAI_10,Mooij_et_al_NIPS_11,Shimizu++2011,Janzing_et_al_AI_12,HyvarinenSmith2013,Peters2014biom,Kpotufe++2014,Nowzohour2015,Sgouritsa++2015}.}

Following the previous rationale, and focusing on the non-deterministic scenario, we here explore the framework of nonlinear additive noise models (ANMs)~\cite{Hoyer08}, which rely on the principle of independence between the cause (prior) and the generating mechanism (conditional).
{A practical algorithmic instantiation of such principle only requires 1) two regression models to learn the two possible forward and backward directions of causation, and 2) the estimation of {\em statistical independence} between the obtained residuals by the factorization and the observation. The direction leading to more independent residuals is decided to be the cause.} 
The method has achieved state-of-the-art results in an exhaustive comparison involving bivariate causal problems in biology, geosciences, economy and social sciences, as illustrated in~\cite{Mooij16jmlr}. 
%\red{The proposed method decides about the causal direction based on the dependence of the residuals obtained after fitting a regression model in the forward and inverse directions, which indicates the true data-generating mechanism.} 
Authors in~\cite{Hoyer08} suggested the use of Gaussian processes~\cite{Rasmussen06,CampsValls16grsm} for regression fitting arbitrarily complex functions, and the Hilbert Schmidt Independence Criterion (HSIC)~\cite{Gretton05} as dependence test based on the excellent converge properties to the true dependence and ease of calculation. HSIC has been previously used in remote sensing for feature selection and dependence estimation~\cite{campsvalls09grsl,Camps-VallsTLM10}. 

We here propose an alternative criterion to the direct dependence of the residuals, and focus on the sensitivity (derivative) of the HSIC dependence estimate. %, \red{as it contains useful information about the asymmetry of the forward and inverse conditional densities.} 
The paper extends our previous work in~\cite{PerezSuay17igarss} with more theoretical insight and a larger set of experimental evidence. In particular, we illustrate performance in a collection of 28 geoscience causal inference problems, in a large database of radiative transfer models simulations and machine learning emulators in vegetation parameter modeling leading to 182 causal problems with ground truth, %in the problem of identifying the atmospheric inversion layer from infrared sounding data,
and in assessing the impact of different regression models in a carbon cycle problem. The criterion achieves state-of-the-art detection rates above chance levels, it is robust to noise sources and distortions, and the adoption of different regression models. 
The presented approach confirms the validity in observational bi-variate problems in the Earth sciences.

The remainder of the paper is organized as follows. 
Section~\ref{sec:causal} reviews the main aspects of the adopted causal framework, and the needed tools for the practical implementation: Gaussian processes for regression and the HSIC estimate for dependence estimation. 
Section~\ref{sec:proposal} derives the HSIC sensitivity maps and describes the proposed causal criterion. 
Section~\ref{sec:results} gives experimental evidence of performance in a wide range of bivariate Earth system problems. 
Finally, we conclude in Section~\ref{sec:conclusions} with some remarks and future outlook.

\section{Causality in pairs of instantaneous variables} \label{sec:causal}

\subsection{Deterministic setting}\label{sec:deterministic}

%\red{Identifying cause and effect in {\em pairs of instantaneous variables} is a very challenging problem. Broadly speaking, given two variables $x$ and $y$, one could ideally identify the cause and effect by analyzing the asymmetry of the corresponding densities $p(x)$ and $p(y)$.}

Let us first consider the case where noise is not present. This deterministic causal problem has been treated before~\cite{Janzing12,Daniusis12,Peters18}, and is often known as Information-Geometric Causal Inference (IGCI). Formally, let us assume a continuous differentiable transformation, $f:x\to y$. The function $f$ has to be a diffeomorphism (it is differentiable and bijective and it has a differentiable inverse) of $[0,1]$ that is strictly monotonic and satisfies $f(0)=0$ and $f(1)=1$. 
Now, using the standard formula of densities under transformation
$$p(x) = p(y)\bigg|\dfrac{\partial f(x)}{\partial x}\bigg|,$$
one could arguably identify the direction of causation from the particular structure of the densities. If the structure of $p(x)$ is not correlated with the slope of $f$, then flat regions of $f$ induce peaks in $p(y)$. The causal hypothesis $y\to x$ is thus implausible because the causal mechanism $f^{-1}$ appears to be adjusted to the `input' distribution $p(y)$. 
Note that here the principle of independence of cause and mechanism reduces to estimate the independence of $p(x)$ and $f$, which interestingly implies dependence between $p(y)$ and $f^{-1}$. %This was the method proposed in~\cite{Daniusis12}. 
This is illustrated in Fig.~\ref{fig:deterministic}. 

We should note, however, that such justifications always refer to oversimplified models that are unlikely to describe realistic situations. After all, bijective deterministic relations are rare in nature. Therefore, IGCI only provides a limited, unrealistic scenario for which cause-effect inference is possible by virtue of an approximate cause-mechanism independence assumption. 

\begin{figure}[t!]
\begin{center}
     \includegraphics[width=.9\columnwidth]{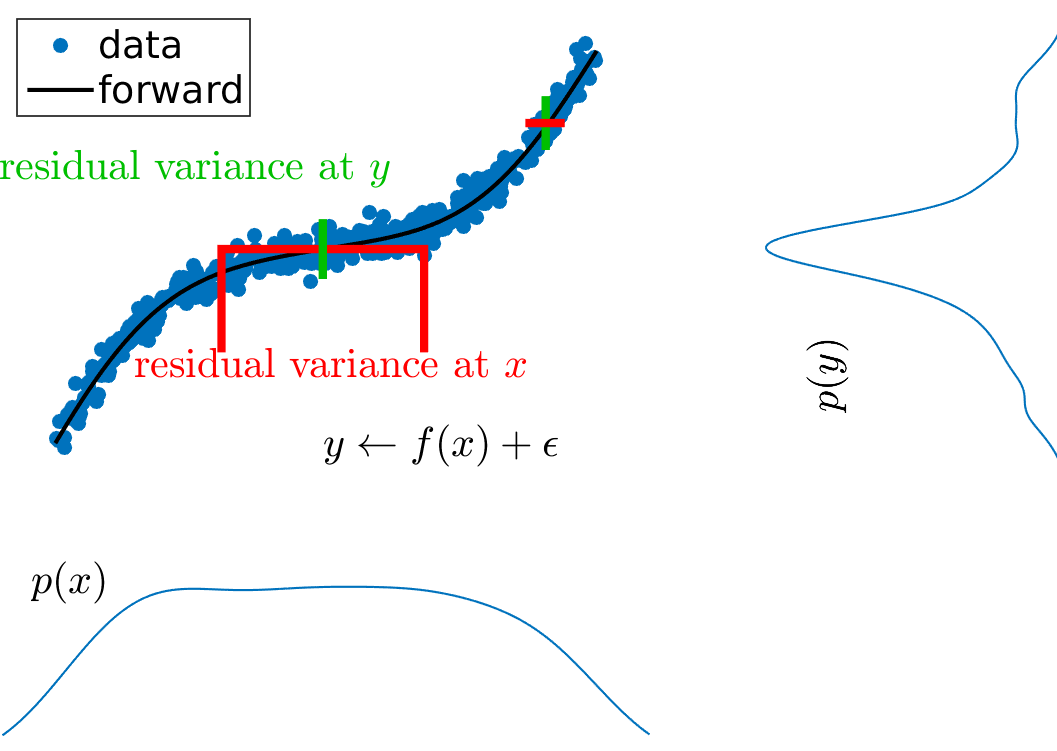}
\end{center}
\caption{\blue{
Cause-effect relations can be identified in a deterministic case by looking at the correlation of the slope of $f$ with the density of the cause, $p(x)$. The rationale is that the cause is generated independently from the mechanism mapping it to the effect. Therefore, the shape (slope) of the function should be more uncorrelated with the density of the cause than the other way around. Otherwise, the inverse $f^{-1}$ seems better fitted to the effect density $p(y)$. The density asymmetry is one of the footprints one can use to infer cause-effect relations.
When noise is present (blue dots), ANM can be applied here: two models are developed (one that tries to explain $x$ from $y$, and vice versa) and then one looks which one yields residuals more uncorrelated with the potential cause. In this example of $x$ causes $y$, one can see that the residual variance of $x$ depends of $y$ (red bars) while the residual variance at $y$ (green bars) is roughly constant for all $x$, which suggests that is a more plausible model.} \label{fig:deterministic} 
%%\caption{Cause-effect relations can be identified in a deterministic case by looking at the correlation of the slope of $f$ with the density of the cause, $p(x)$. The shape (slope) of the function is incorrelated with the density of the cause. Otherwise, the inverse $f^{-1}$ seems better fitted to the effect density $p(y)$. \label{fig:deterministic} 
}
\end{figure}
\iffalse
\begin{figure}[t!]
\centerline{\includegraphics[width=9.5cm]{images/deterministic.pdf}}
\caption{Cause-effect relations can be identified in a deterministic case by looking at the correlation of the slope of $f$ with the density of the cause, $p(x)$. The rationale is that the cause is generated independently from the mechanism mapping it to the effect. Therefore, the shape (slope) of the function should be more uncorrelated with the density of the cause than the other way around. Otherwise, the inverse $f^{-1}$ seems better fitted to the effect density $p(y)$. The density asymmetry is one of the footprints one can use to infer cause-effect relations.
When noise is present (blue dots), ANM can be applied here: two models are developed (one that tries to explain $x$ from $y$, and vice versa) and then one looks which one yields residuals more uncorrelated with the potential cause. In this example of $x$ causes $y$, one can see that the residual variance of $x$ depends of $y$ (red bars) while the residual variance at $y$ (green bars) is roughly constant for all $y$, which suggests that is a more plausible model.
\label{fig:deterministic} 
}
\end{figure}
%\red{Mirar \url{https://www.youtube.com/watch?v=hMiqgKqu8O8}}
\fi

\subsection{Non-deterministic setting} \label{sec:stochastic}

The vast majority of real (and thus more interesting) problems are not deterministic, the function $f$ relating the two variables is not bijective, and $f$ is inaccessible so one cannot compute its derivatives, and from there a criterion of independence between the densities of the cause and the mechanism.
An alternative is to rely on the assumption made in Additive Noise Models (ANM) for causal inference~\cite{Hoyer08}. 

Given two random variables $x$ and $y$ with causal relation $x\to y$, under some conditions, one can demonstrate (cf. e.g. Theorem 4.5 about the identifiability of ANMs in~\cite{Peters18}) that there exists an additive noise model
$$y = f(x) + n_f,$$
in the correct causal direction, but there exists no additive noise model 
$$x = g(y) + n_g$$
in the anticausal direction. \glue{The above mentioned Theorem can be derived through the particular assumptions and it states that generically, a distribution does not admit an ANM in both directions at the same time.}

This observation allowed Hoyer et al.~\cite{Hoyer08} to design an algorithm to distinguish cause from effect in pairs of variables from empirical data. Essentially, the methodology performs nonlinear regression from $x\to y$ (and vice versa, $y\to x$) and assesses the independence of the forward, $r_f=y-f(x)$, and backward residuals, $r_b = x-g(y)$, with the input variables (drivers) $y$ and $x$, respectively. The more independent residuals tell us the right direction of causation.%%\red{The statistical significance of the independence test tells the right direction of causation. }

\blue{Relevant assumptions are done in this approach. First, we assume that the considered problem is described accurately looking at the pairs ({\em representational property}). Another important assumption is the {\em causal sufficiency}, which states that there is no hidden
common cause in the considered variables that is causing any of the latter, and thus acts as a confounder. Third, the {\em causal Markov assumption} hold, which allows us to treat the causal graph as a probabilistic one. In addition, there are two extra conditions on the regression functions: (i) $f,g$ are either linear and noise is non-Gaussian, or (ii) $f,g$ are nonlinear and output's densities are positive and smooth~\cite{Mooij16jmlr,Peters18}.}

%Note that in a deterministic noise-free case, cf.~\S\ref{sec:deterministic}, we cannot compute the independence between noise and the cause to find the causal direction. Nevertheless, in such cases, one still may exploit a certain type of independence  between the transformation $f$ and the distribution of the cause $x$ to characterize the causal asymmetry and determine the causal direction. This observation links the deterministic and stochastic approaches.

In order to define a practical criterion, two ingredients are needed only: a regression method to learn functions $f$ and $g$, and a powerful dependence estimate to assess the independence of the residuals. Therefore, the framework needs two fundamental blocks: 1) a nonlinear regression model, and 2) a dependence measure. We typically rely on Gaussian Processes~\cite{CampsValls16grsm} and the HSIC~\cite{Gretton05}, respectively. In what follows, we briefly review the theory under these two kernel methods~\cite{CampsValls09wiley}. Then we define the causality criterion.

\subsubsection{Gaussian Processes (GPs)}

Standard regression approximates observations (often referred to as \emph{outputs}) as the sum of some unknown latent function $f(\x)$ of the input data plus constant power (homoscedastic) Gaussian noise of variance $\lambda^2$. Note that in our case, both inputs and outputs are unidimensional, $x,y\in\Real$. Therefore, given $n$ input-output data pairs, the dataset is denoted as ${\mathcal D} = \{(\x_i,\y_i)\}_{i=1}^{n}$, and the model approximation is
\begin{equation}\label{GLR}
y_i = f(\x_i) + \varepsilon_i,~~~\varepsilon_i \sim\Normal(0,\lambda^2).
\end{equation}
Instead of proposing a parametric form for $f(\x)$ and learning its parameters in order to fit observed data well,  GP regression proceeds in a Bayesian, non-parametric way. A zero mean\footnote{It is customary to subtract the sample mean to data $\{y_i\}_{i=1}^n$, and then to assume a zero mean model.} GP prior is placed on the latent function $f(\x)$ and a Gaussian prior is used for each latent noise term $\varepsilon_i$, 
$f(\x)\;\sim\;\GP(0, k_\vect{\theta}(\x,\x'))$, 
where $k_\vect{\theta}(\x,\x')$ is a covariance function parametrized by $\vect{\theta}$ and $\sigma^2$ is a hyperparameter that specifies the noise power.
Essentially, a GP is a stochastic process whose marginals are distributed as a multivariate Gaussian. In particular, given the priors $\GP$, samples drawn from $f(\x)$ at the set of locations $\{\x_i\}_{i=1}^n$ follow a joint multivariate Gaussian with zero mean and covariance matrix $\mat{K_\vect{ff}}$ with  $[\mat{K_\vect{ff}}]_{ij} = k_\vect{\theta}(\x_i,\x_j)$.

If we consider a test location $\x_*$ with corresponding output $y_*$, priors $\GP$ induce a %the following  joint 
prior distribution between the observations ${\bf y}=[y_1,\ldots,y_n]^\top$ and $y_*$. 
%Collecting available data in $\dataset\equiv\{\x_n,y_n|n=1,\ldots N\}$, 
Now it is possible to analytically compute the posterior distribution over the unknown output $y_*$ given the test input $\x_*$ and the available training set $\dataset$, 
\begin{eqnarray}
 \prob(y_*|\x_*,\dataset) = \Normal(y_*|\mu_{\text{GP}*},\sigma_{\text{GP}*}^2),
\end{eqnarray}
which is a Gaussian with the following mean and variance:
\begin{align}
\mu_{\text{GP}*} &= \vect{k}_{\vect{f}*}^\top (\mat{K}_{\vect{ff}}+\lambda^2\mat{I}_n)^{-1}\vect{
y}  \label{eq:mugp}\\
\sigma_{\text{GP}*}^2 &= \lambda^2+k_{**}- \vect{k}_{\vect{f}*}^\top (\mat{K}_{\vect{ff}}+\lambda^2\mat{I}_n)^{-1}\vect{k}_{\vect {f}*}, \label{eq:sigmagp}
\end{align}
where $\vect{k}_{\vect{f}*} = [k(\x_*,\x_1),\ldots,k(\x_*,\x_n)]^\top\in\Real^{n}$ contains the kernel similarities of the test point $\x_*$ to all training points in $\dataset$, $\mat{K}_{\vect{ff}}$ is a $n\times n$ kernel (covariance) matrix whose entries contain the similarities between all training points, %${\bf y} = [y_1,\ldots,y_N]^\top\in\Real^{N\times 1}$, 
$\lambda^2$ is a hyperparameter accounting for the variance of the noise, $k_{**}=k(\x_*,\x_*)$ is a scalar with the self-similarity of $\x_*$, and $\mat{I}_n$ is the identity matrix of size $n$. Note that both the predictive mean and the variance can be computed in closed-form, and that the predictive variance $\sigma_{\text{GP}*}^2$ does not depend on the outputs/target variable. %Also note that the predictive mean is computable in $\bigO(N^3)$ time since it involves the inversion of the $N\times N$ matrix ($\mat{K}_{\vect{ff}}+\sigma^2\mat{I}$), see \cite{Rasmussen06}.}
%Note that in our case, we will limit to a one-dimensional fitting so both $\x$ and $y$ are scalars.

\subsubsection{Kernel Dependence Estimation with HSIC}

Let us consider two spaces ${\mathcal X}, {\mathcal Y}\subseteq \Real$, on which we jointly sample observation pairs $(\x,\y)$ from distribution $p(\x,\y)$. The covariance matrix ${\mathcal C}_{\x\y}$ encodes first order dependencies between the random variables. A statistic that efficiently summarizes the content of this matrix is its Hilbert-Schmidt norm. This quantity is zero if and only if there exists no second order dependence between $\x$ and $\y$.

The nonlinear extension of the notion of covariance was proposed in \cite{Gretton05} \glue{to cope with higher-order relations between the data. The use of its linear formulation has some limitations and cannot discover more complex relations, for this purposed the use of nonlinear kernel functions allows to capture higher-order effects.} Let us define a (possibly non-linear) mapping $\boldsymbol{\phi}$: ${\mathcal X}\to {\mathcal F}$ such that the inner product between features is given by a positive definite (p.d.) kernel function  $K_x(\x,\x') = \langle \boldsymbol{\phi}(\x),\boldsymbol {\phi}(\x')\rangle$. The feature space ${\mathcal F}$ has the structure of a reproducing kernel Hilbert space (RKHS). Let us now denote another feature map $\boldsymbol{\psi}$: ${\mathcal Y}\to {\mathcal G}$ with associated p.d. kernel function $K_y(\y,\y') = \langle \boldsymbol{\psi}(\y),\boldsymbol{\psi}(\y')\rangle$. Then, the cross-covariance operator between these feature maps is a linear operator ${\mathcal C}_{\x\y}:{\mathcal G} \to {\mathcal F}$ such that
$
{\mathcal C}_{\x\y} = 
%{\mathbb E}_{\x\y}[(\boldsymbol{\phi}(\x)-\mu_x)(\boldsymbol{\psi}(\y)-\mu_y)^\top],
{\mathbb E}_{\x\y}[(\boldsymbol{\phi}(\x)-\boldsymbol{\mu}_x)\otimes(\boldsymbol{\psi}(\y)-\boldsymbol{\mu}_y)],
$
where $\otimes$ is the tensor product, $\boldsymbol{\mu}_x={\mathbb E}_{\x}[\boldsymbol{\phi}(\x)]$, and $\boldsymbol{\mu}_y={\mathbb E}_{\y}[\boldsymbol{\psi}(\y)]$. 
%, and ${\bf u}{\bf v}^\top$ denotes the linear operator ${\bf u}{\bf v}^\top: {\mathcal G} \to {\mathcal F}$, ${\bf w}\mapsto {\bf u} \langle {\bf v},{\bf w}\rangle$. 
See more details in \cite{Baker73,Fukumizu04}. The squared norm of the cross-covariance operator, $\|{\mathcal C}_{\x\y}\|^2_{\text{HS}}$, is called the Hilbert-Schmidt Independence Criterion (HSIC) and can be expressed in terms of kernels~\cite{Gretton05}.
%\begin{eqnarray}
%\begin{array}{lll}
%\text{HSIC}({\mathcal F},{\mathcal G},{\mathbb P}_{\x\y}) & 
%= & \|{\mathcal C}_{\x\y}\|^2_{\text{HS}} \nonumber\\[2mm]
%& = &{\mathbb E}_{\x\x'\y\y'}[k_x(\x,\x')k_y(\y,\y')] \\[2mm]
%& & + {\mathbb E}_{\x\x'}[k_x(\x,\x')]{\mathbb E}_{\y\y'}[k_y(\y,\y')]  \nonumber\\ [2mm]
%& & - 2{\mathbb E}_{\x\y}[{\mathbb E}_{\x'}[k_x(\x,\x')] {\mathbb E}_{\y'}[k_y(\y,\y')]], \nonumber
%\end{array}
%\end{eqnarray}
%where ${\mathbb E}_{\x\x'\y\y'}$ is the expectation over both $(\x,\y) \sim {\mathbb P}_{\x\y}$ and an additional pair of variables $(\x',\y') \sim {\mathbb P}_{\x\y}$ drawn independently according to the same law.
%An empirical estimate of HSIC can be easily obtained. 
%the sample data pairs ${\bf x}, {\bf y} \in\Real^{n}$, 
%and the joint dataset ${\mathcal Z}={\mathcal X}\times {\mathcal Y}$, 
Given the set ${\mathcal{D}}$ with $n$ scalar pairs drawn from the joint $p(\x,\y)$ %%$n$ scalar pairs $\{(\x_1,\y_1),\ldots,(\x_n,\y_n)\}$ 
%collectivelly grouped in a sample matrix ${\bf Z}\in\Real^{n\times(d_x+d_y)}$ 
%drawn from the joint ${\mathbb P}_{\x\y}$, 
an empirical estimator of HSIC is~\cite{GreBouSmoSch05}:
\begin{eqnarray}
\text{HSIC}= %\dfrac{1}{n^2} \text{Tr}({\bf K}_x{\bf H}{\bf K}_y{\bf H}) = 
\dfrac{1}{n^2} \text{Tr}({\bf H}{\bf K}_x{\bf H}~{\bf K}_y),
\label{empHSIC}
\end{eqnarray}
where $\text{Tr}(\cdot)$ is the trace operation (the sum of the diagonal entries), ${\bf K}_x$, ${\bf K}_y$ are the kernel matrices for the input random variables $\x$ and $\y$, respectively, and ${\bf H}= {\bf I} - \frac{1}{n}\mathbbm{1}\mathbbm{1}^\top$ centers the data in the feature spaces ${\mathcal F}$ and ${\mathcal G}$, respectively. %Here $\delta$ represents the Kronecker symbol, where $\delta_{i,j}=1$ if $i=j$, and zero otherwise. 
It is important to note that HSIC$=0$ occurs if and only if $\x$ and $\y$ are \glue{independent, the proof of this theoretical result is provided on~\cite{GreBouSmoSch05}}.

\subsubsection{Causal criterion}

%\red{Depending on the significance levels for rejecting and accepting independence, one may get an ANM in both directions, in no direction, or in one direction. To enforce decisions, one just infers the direction to be the causal one, for which the $p$-value for rejecting independence is higher.}

We build on the ANM approach revised in Section~\ref{sec:stochastic} and originally presented in~\cite{Hoyer08} to discover causal association between variables $x$ and $y$. The method provided good results in a set of experiments. A thorough comparison to other methods and in many real and synthetic datasets was conducted in~\cite{Mooij16jmlr}. In general, the best performing criterion to detect the causal direction was simply defined as the difference in test statistic between both forward and backward models:
\begin{eqnarray}\label{eq:C}
\widehat C:=\text{HSIC}(x,r_f)-\text{HSIC}(y,r_b)
\end{eqnarray}
where $r_f=y-f(x)$ and $r_b=x-g(y)$ are the residuals yielded by the forward (backward) models $f$ and $g$, respectively. See Section~\ref{sec:stochastic}. 
Intuitively, we compare which method yields more independent residuals (lower HSIC value) as an indicator of model plausibility.
The sign of the criterion $\hat C$ tells the causal direction: more negative values indicate that the forward model is more plausible and thus one decides that $x$ causes $y$. While other criteria could be adopted, we take this one as the baseline method because of its state-of-the-art performance~\cite{Mooij16jmlr}.

\subsection{An illustrative example}\label{sec:examplehoyer}

\begin{figure}[h!]
\setlength{\tabcolsep}{-2pt}
\begin{tabular}{cc}
\includegraphics[width=4.5cm]{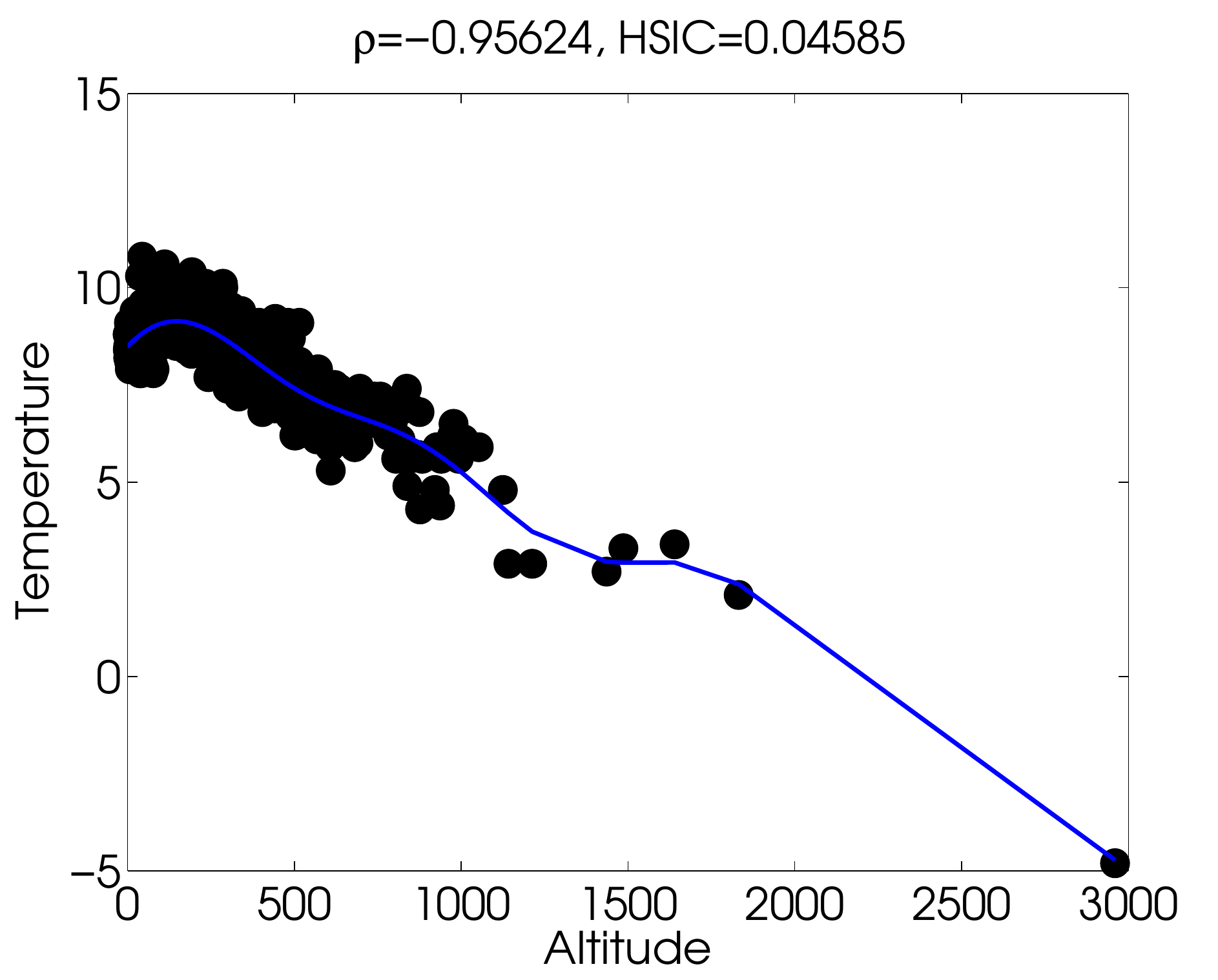} & \includegraphics[width=4.5cm]{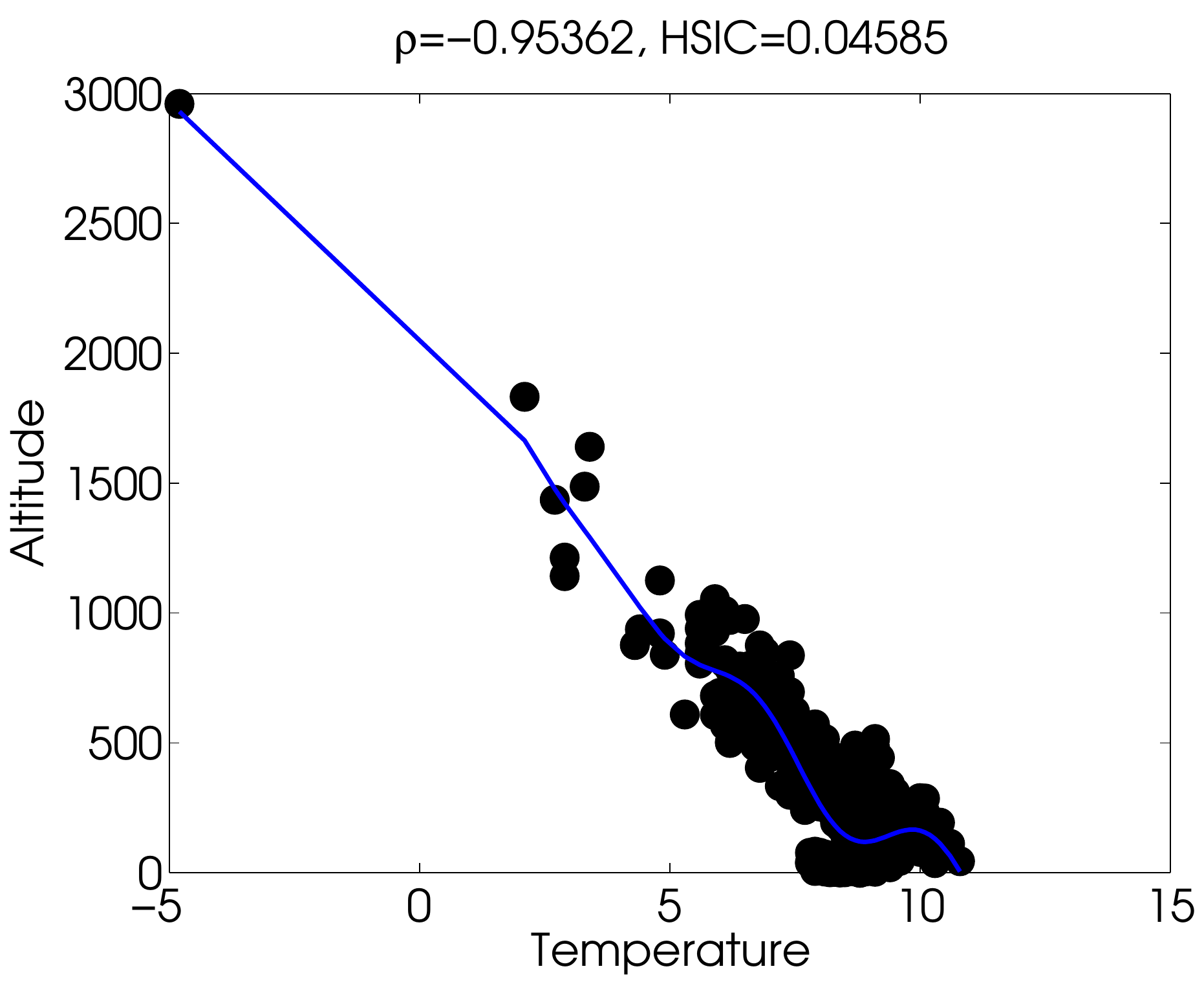} \\
\includegraphics[width=4.5cm]{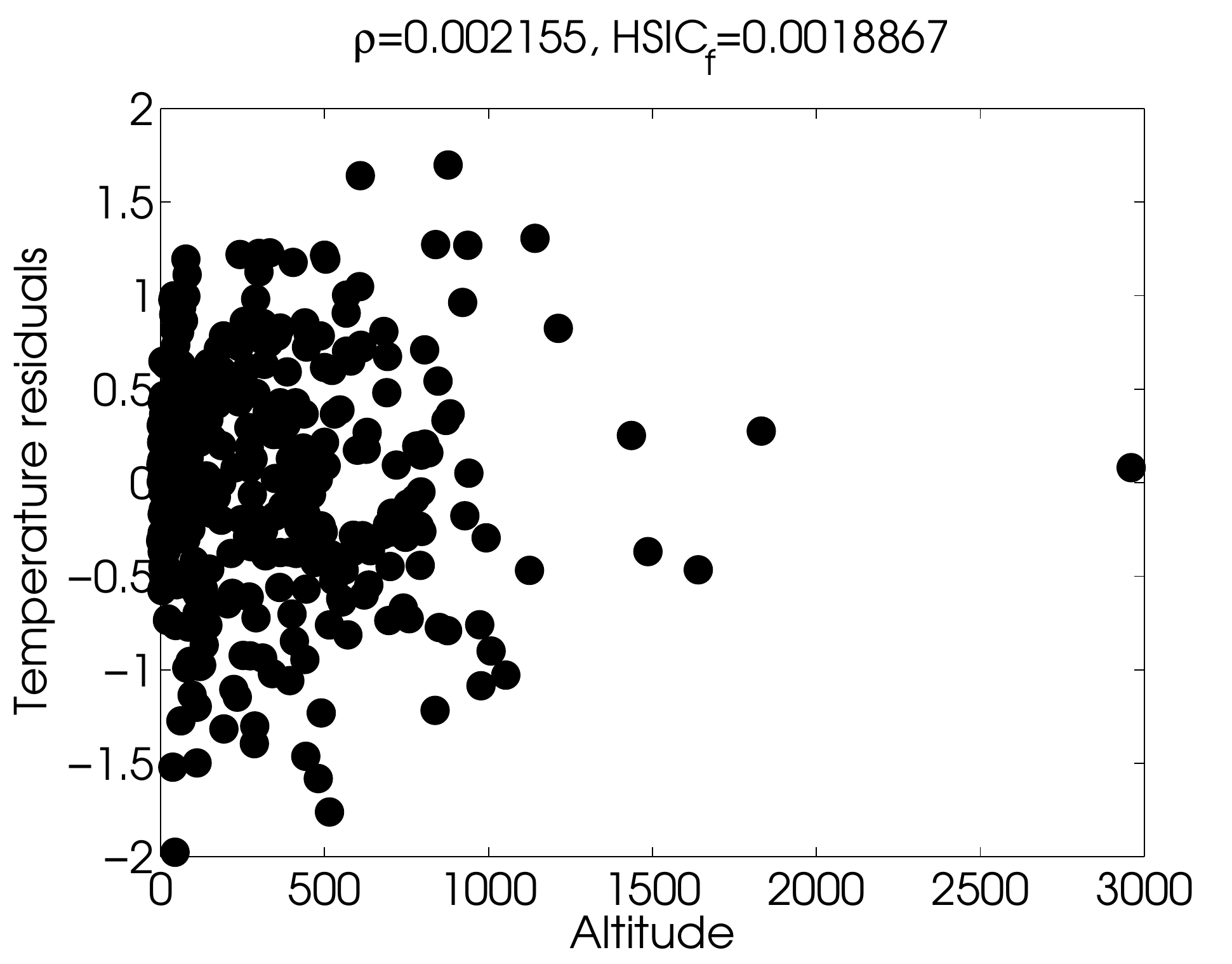} & \includegraphics[width=4.5cm]{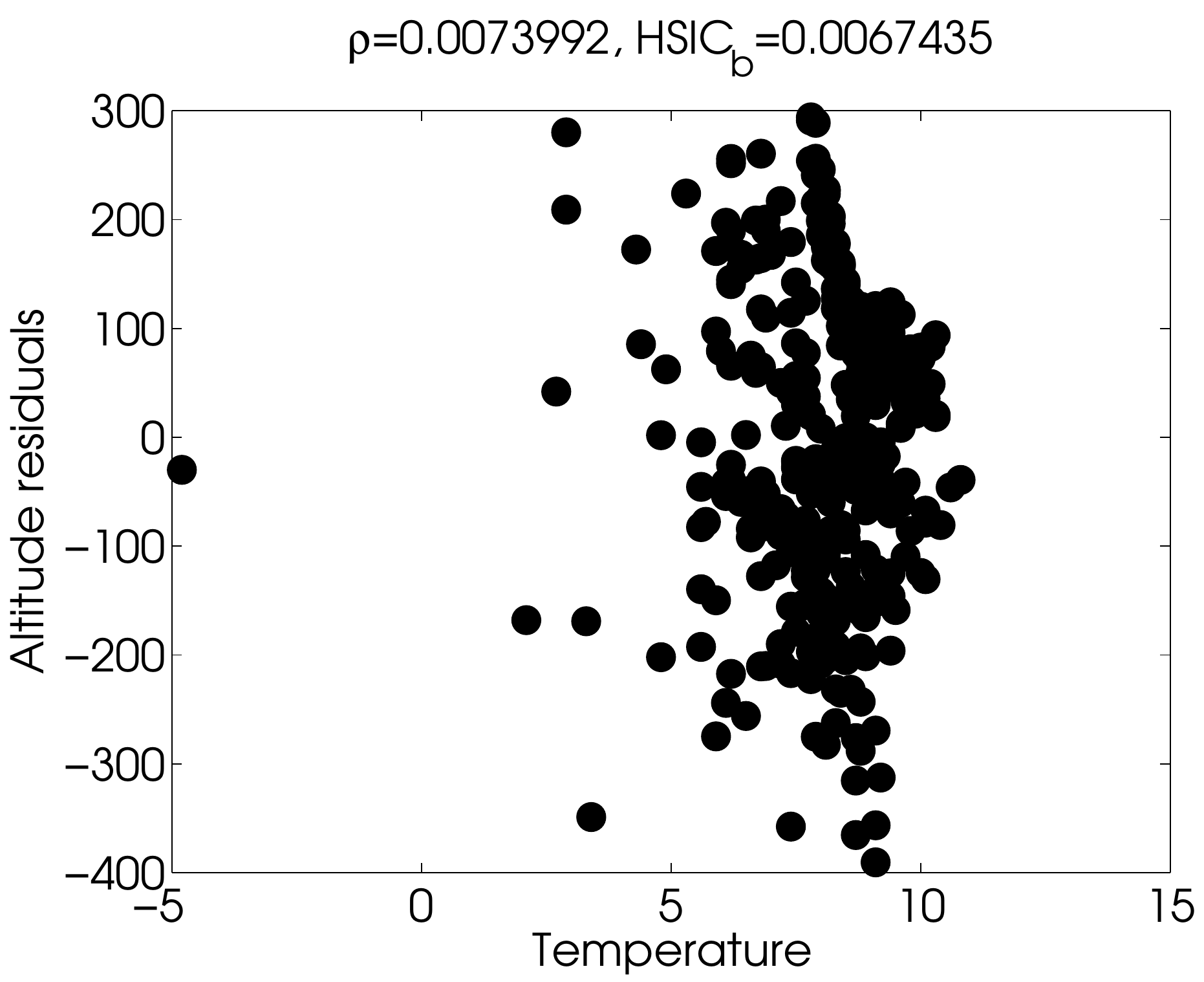} \\
\end{tabular}
\caption{Example of the method in~\cite{Hoyer08} in the altitude causes temperature problem. The fitted functions (top row) are used to approximate $f$ and $g$ models. From here, the particular distribution of the obtained residuals \blue{versus the potential cause} (bottom) establish the direction of causation as that showing more independent residuals from the driver. We give the correlation coefficient and the HSIC values \blue{between the residuals and the potential cause under examination for both the forward and backward models, i.e. between residuals $r_f=y-\hat y$ and $x$, and between residuals $r_b=x-\hat x$ and $y$. In this work we focus on HSIC to capture the nonlinear dependence.}\label{fig:stochastic}}
\end{figure}

The intuition behind the approach is that statistically significant residuals in one direction indicates the true data-generating mechanism (see Sections~\S\ref{sec:intro} and \S\ref{sec:deterministic}). 
See an illustrative example in Fig.~\ref{fig:stochastic}. The problem contains values of altitude and averaged temperature of $n=349$ weather stations in Germany, and the data was provided by the Deutscher Wetterdienst (DWD). The problem reduces to identify the common sense direction of `altitude causes temperature' from the data. Cities are obviously in the troposphere, so under the inversion layer. Nevertheless, a potential confounder is the latitude, since in Germany most of the mountains are in the south, which leads to positive correlations between altitude and temperature. Nevertheless, the direct causal relation between altitude and temperature dominates over the confounder. 
Following Hoyer's approach, two GPs were fitted to the forward and backward directions, and we measured the independence of the obtained residuals with both the standard correlation coefficient $\rho$ and the HSIC. Lower independence values are obtained in the forward (causal) direction. % (HSIC$_f<$HSIC$_b$).

\section{Causal Inference with Sensitivity Maps} \label{sec:proposal}

%HSIC has demonstrated excellent capabilities to detect independence between random variables but, as for any kernel method, the learned relations are hidden behind the kernel feature mapping. %Visualization and geometrical interpretation of kernel methods in general and kernel dependence estimates in particular is an important issue in machine learning. 
%To address this issue, we next derive sensitivity maps for HSIC that account for the relevance of the examples in the dependence measure. This motivates the introduction of the sensitivity maps of the measure as a novel criterion for causal inference. We hypothesize that the most sensitive points in the dependence measure dominate the causal discrimination.

\blue{HSIC has been used in combination with ANM for causal inference before~\cite{Mooij16}, see Eq.~\eqref{eq:C}. Since the most sensitive points typically dominate the HSIC measure of dependence, we here propose a criterion for causal discrimination in terms of the derivative of the HSIC with respect both the drivers and the residuals in ANM schemes. In the following, we review the main ingredients of the proposed criterion: we give the formulation of the HSIC sensitivity maps, the proposed causality criterion, and study their empirical and theoretical properties. }

\subsection{Sensitivity analysis and maps}

A general definition of the {\em sensitivity map} (SM) in the context of kernel methods was originally introduced in~\cite{Kjems02}. Let us define HSIC as a function $h:\mathcal{X}\times \mathcal{Y}\to\Real$. The sensitivity map is the expected value of the squared derivative of the function (or the log of the function) with respect the arguments $\param=(x,y)$. Formally, let us define the sensitivity as 
\begin{equation}
s = \int_{\mathcal Z}\bigg(\dfrac{\partial h(\param)}{\partial z}\bigg)^2p(\param)\text{d}\param,
\label{smeander}
\end{equation}
where $p(\param)$ is the probability density function over the inputs $\param\in{\mathcal Z}$. Intuitively, the objective of the sensitivity map is to measure the changes of the function $h(\param)$ along the inputs $\param$. 
In order to avoid the possibility of cancellation of the terms due to its signs, the derivatives are typically squared, even though other unambiguous transformations like the absolute value could be equally applied. Integration gives an overall measure of sensitivity over the observation space ${\mathcal Z}$. The {\em empirical sensitivity map} approximation to Eq.~\eqref{smeander} is obtained by replacing the integral with a summation over the available $n$ samples 
\begin{equation}\label{empirical}
s\approx\dfrac{1}{n}\sum_{i=1}^n \dfrac{\partial h(\param)}{\partial z}\bigg|_{\param_i}^{~2}.
\end{equation}
Since the HSIC function $h$ works on $(x,y)$, we have to apply this expression twice, which returns the sensitivity vector $\s=[s^x,s^y]$.  This gives the relevance of variables $x$ and $y$ in the function $h$. 
Alternatively, one can summarize the maps variable-wise to obtain a point-wise relevance:
\begin{equation}\label{empirical2}
p_i = \sqrt{\bigg(\dfrac{\partial h(\param)}{\partial x}\bigg)^{2} + \bigg(\dfrac{\partial h(\param)}{\partial y}\bigg)^{2}},
\end{equation}
which will be used in this paper to evaluate the relevance of points in the dependence measure. Actually, this way of summarizing the information conveyed by the sensitivity map is somewhat related to the concepts of {\em leveraging} and {\em influential points} in statistics~\cite{burt2009elementary}.

\subsection{Sensitivity maps for the HSIC}

In order to derive the sensitivity maps for HSIC which were originally presented in~\cite{PEREZSUAY2017}, we need to compute its partial derivatives w.r.t. points in variables $x$ and $y$, i.e. $x_{i}$ and $y_{i}$. By applying the chain rule, and first-order derivatives of matrices, we obtain:
%Now, deriving the HSIC estimate in Eq.~\eqref{empHSIC} w.r.t. every $i$-th example and $j$-th feature of $\X$ and $\Y$, it is easy to find: %$${\bf s}_j^x:=\dfrac{\partial\text{HSIC}}{\partial{\bf x}^r}=-\frac{1}{\sigma^2(n-1)^2}{\bf K}_x{\bf H}{\bf K}_y{\bf H}\sum_{k=1}^d\delta_r^k(x_i^k-x_j^k),$$ 
\begin{equation}\label{shsic_x}
%S_{ij}^x:=\dfrac{\partial\text{HSIC}}{\partial X_{ij}}=-\frac{2}{\sigma^2 n^2}\text{Tr}\left({\bf H}{\bf K}_y{\bf H}({\bf K}_x\circ {\bf M}_{ik}^j)\right),
s_{i}^x:=\dfrac{\partial\text{HSIC}}{\partial x_{i}}=-\frac{2}{\sigma^2 n^2}\text{Tr}\left({\bf H}{\bf K}_y{\bf H}({\bf K}_x\circ {\bf M}_i\right)),
\end{equation} 
\glue{where matrix ${\bf M}_i$ depends on the $i$-th point sample and it is formed by zeroes except in the $i$-th column which corresponds to $\text{vec}(x_i)-\x$ ($i$-th element minus vector $\x$).} %This expression can be used to better implement the sensitivity without computing the whole $M_i$ matrix, just its $i$-th column.}
%%\blue{where matrix ${\bf M}_i$ is full of 0 except in the $i$-th column which corresponds to $\text{vec}(x_i)-\x$ ($i$-th element minus vector $\x$). This expression can be used to better implement the sensitivity without computing the whole $M_i$ matrix, just its $i$-th column.}
\begin{equation}\label{shsic_y}
s_{i}^y:=\dfrac{\partial\text{HSIC}}{\partial y_{i}} = -\frac{2}{\sigma^2 n^2}\text{Tr}\left({\bf H}{\bf K}_x{\bf H}({\bf K}_y\circ {\bf N}_i)\right),
\end{equation}
\glue{where matrix ${\bf N}_i$ depends on the $i$-th point sample and it is formed by zeroes except in the $i$-th column which corresponds to $\text{vec}(y_i)-\y$ ($i$-th element minus vector $\y$).}
%%\blue{where matrix ${\bf N}_i$ is full of 0 except in the $i$-th column which corresponds to $\text{vec}(y_i)-\y$ ($i$-th element minus vector $\y$). This expression can be used to better implement the sensitivity without computing the whole $N$ matrix, just its $i$-th column.}
The joint sensitivity map is defined as the concatenation of individual sensitivities, $\s=[s^x,s^y]$, where $s^x=\frac{1}{n}\sum_{i=1}^n (s_i^x)^2$ and $s^y=\frac{1}{n}\sum_{i=1}^n (s_i^y)^2$; 
and the point-wise sensitivity as $p_i = \sqrt{(s_i^x)^2 + (s_i^y)^2}$.

%It is worth noting that, even though one could be tempted to use each sensitivity map independently, the solution is a {\em vector field}, and we should treat the sensitivity map jointly. \red{To do this we define the {\em total sensitivity map} for all features and samples as ${\bf S}:=[{\bf S}^x,{\bf S}^y]\in\Real^{n\times (d_x+d_y)}$. From ${\bf S}$ we can compute the {\em empirical sensitivity map} per either point or feature by integration in the corresponding domain, whose empirical estimates are respectively $s_i\approx\frac{1}{d}\sum_{j=1}^d S_{ij}^2$ and $s_j\approx\frac{1}{n}\sum_{i=1}^n S_{ij}^2.$}
%\begin{equation}\label{empirical2}
%s_i\approx\dfrac{1}{d}\sum_{j=1}^d \dfrac{\partial \text{HSIC}}{\partial Z_{ij}}\bigg|_{\Param_j}^{~2},~~\text{and}~~
%s_j\approx\dfrac{1}{n}\sum_{i=1}^n \dfrac{\partial \text{HSIC}}{\partial Z_{ij}}\bigg|_{\Param_i}^{~2}.
%\end{equation}
%This complementary view of the sensitivity reports information about the directions most impacting the dependence estimate, and allows a quantitative geometric evaluation of the measure. 

\subsection{Proposed causal criterion with sensitivity maps}

\begin{figure}[h!]
\setlength{\tabcolsep}{-2pt}
\begin{tabular}{cc}
\includegraphics[width=4.5cm]{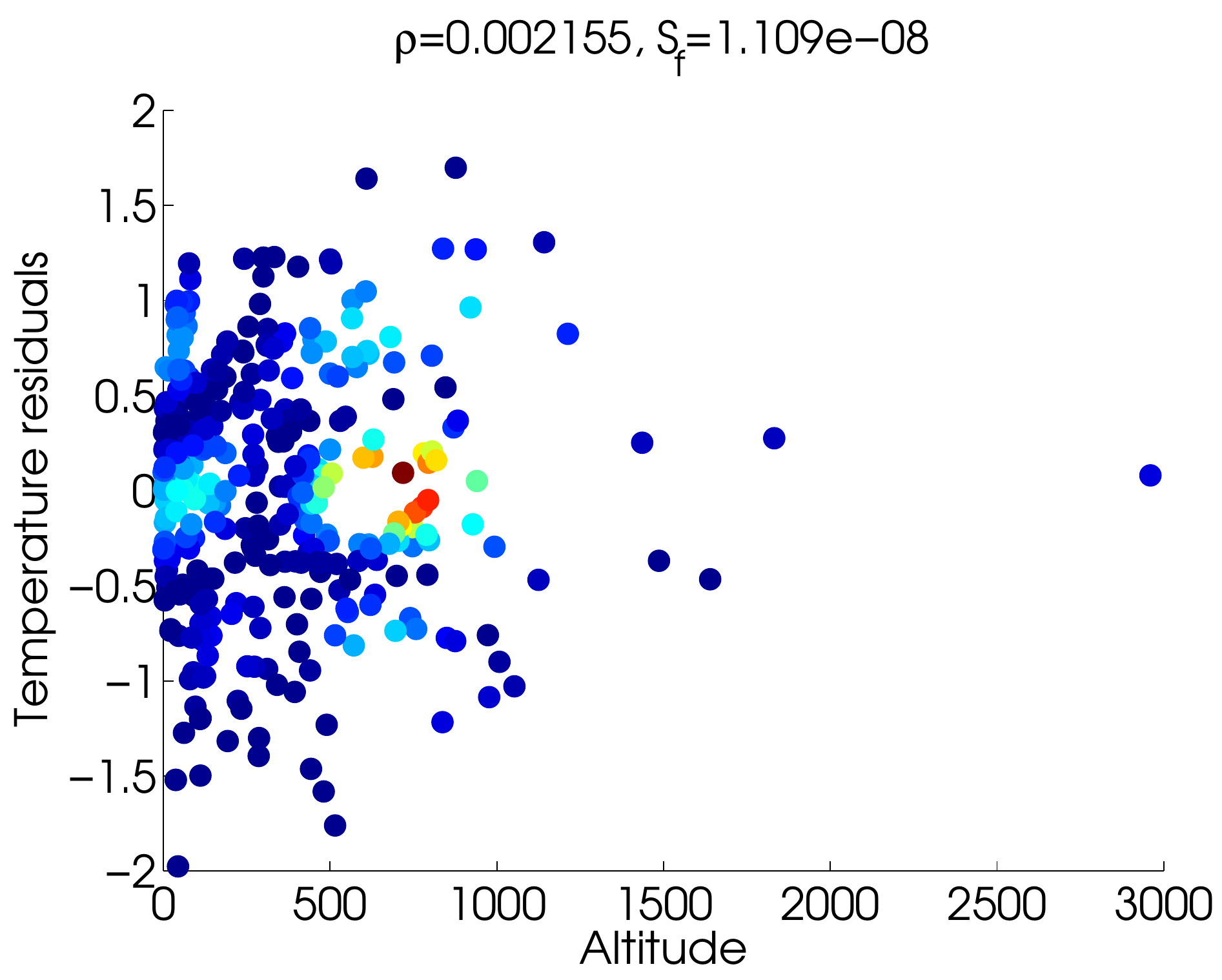} & \includegraphics[width=4.5cm]{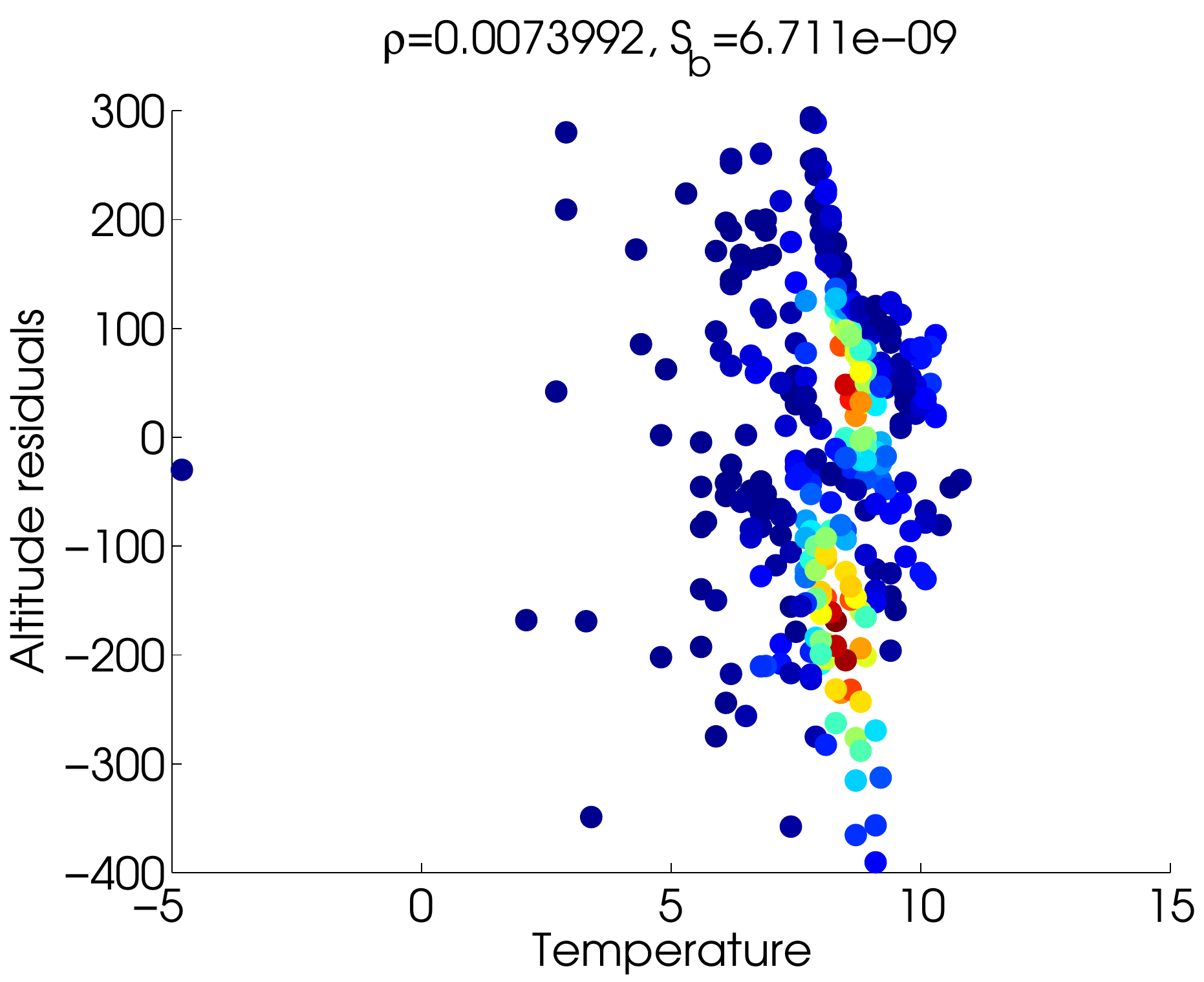} \\
\end{tabular}
\caption{Sensitivity maps of the points most affecting the independence measure (follow-up from example in section~\ref{sec:examplehoyer}). \blue{Colors reflect the importance of the particular example in the dependence measure (HSIC), as computed with the sensitivity map.} \label{fig:sensicriterion}}
\end{figure}

We here propose an alternative criterion for bivariate causality based on the sensitivity maps of HSIC in both directions:
\begin{eqnarray}\label{eq:Cs}
\widehat C_s:=(s_b^y + s_b^{r})-(s_f^x + s_f^{r}),
\end{eqnarray}
where subscripts $f$ and $b$ stand for the forward and backward directions respectively, and the superscripts refer to the sensitivities of either the observations $x$ and $y$, or the corresponding residuals, $r_f$ and $r_b$. The criterion now accounts for the relative relevance of points and residuals in the dependence estimate according to the sensitivity measure. Besides, note the intuitive connection to the deterministic case in section~\ref{sec:deterministic}. Here we do not focus on the derivative of the underlying function, but the derivative of the dependence estimate itself~\cite{Daniusis12}. 
%\red{The sensitivity maps derived from both the forward and backward models allow us to summarize the impact of the examples in the dependence measure.}
Following the same {\em ``altitude causes temperature''} example, we show in Fig.~\ref{fig:sensicriterion} the point sensitivities  in the distributions. It becomes clearer the structured (more dependent) distribution in the backward direction (Fig.~\ref{fig:sensicriterion}[right]), which suggests that the forward direction is the causal one. 

%%%%%%%%%%%%%%%%%

\begin{figure*}[t!]
\begin{center}
\begin{tabular}{ccc}
(a) & (b) & (c)\\
\includegraphics[height=4.5cm]{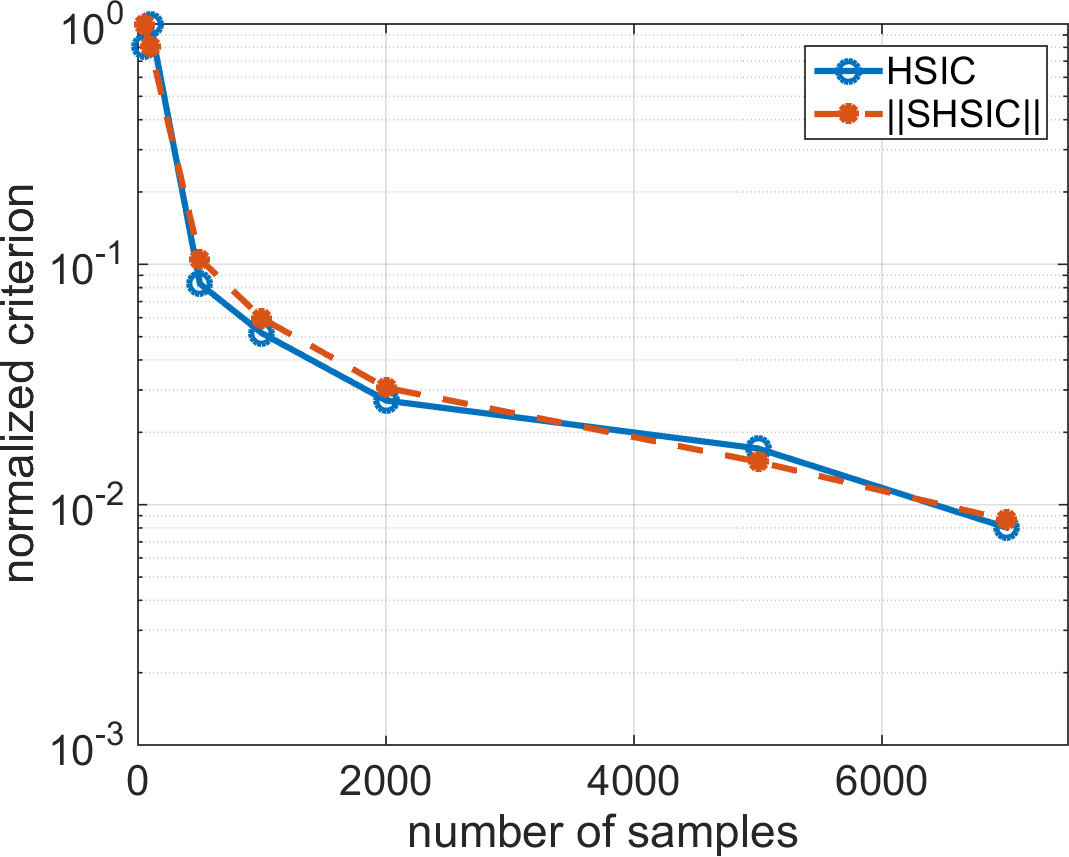} &
\includegraphics[height=4.5cm]{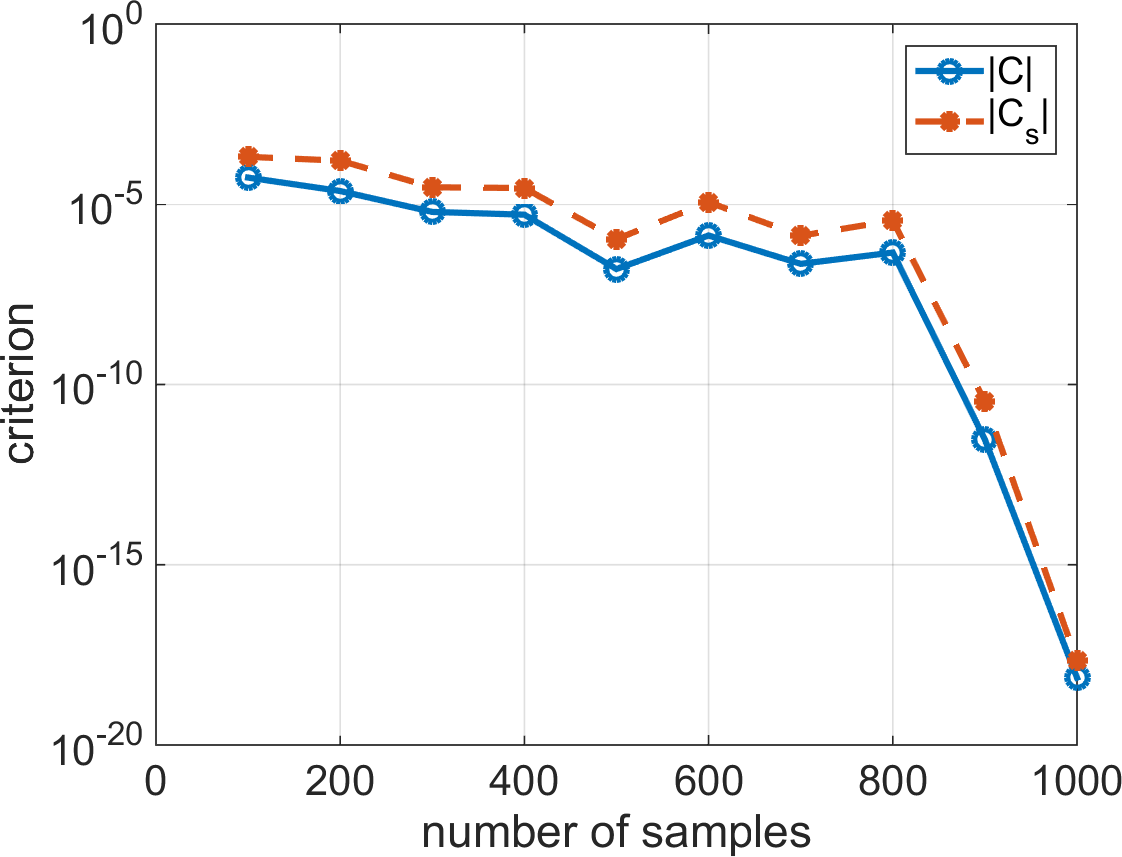}&
\includegraphics[height=4.5cm]{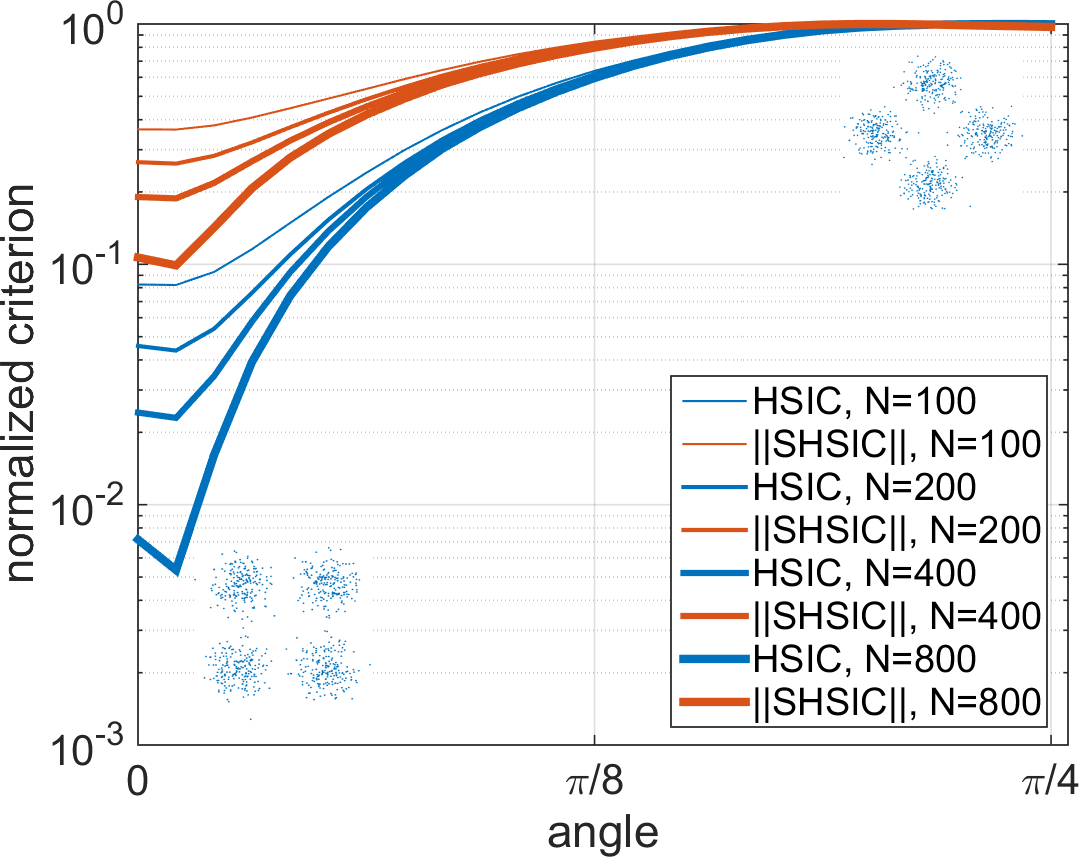}
\end{tabular}
\end{center}
\vspace{-0.25cm}
\caption{\blue{Convergence of (a) the HSIC and the norm of its sensitivity map; (b) the criteria for causal detection, both under the independence hypothesis. Data are drawn from uniform distributions independently, $x,y\sim{\mathcal U}(0,1)$ with different number of samples. Both estimators converge to zero with the number of samples at similar rates. (c) Convergence of the criteria under the hypothesis of varying dependence. We generate different problems from independence (left) to dependence (right) by rotating the X-OR gate for different angles, $\theta\in$[$0,\pi/4$]. We show the HSIC and the norm of the HSIC sensitivity maps when varying the $\theta$ angle. }}
\label{fig:1}
\end{figure*}

\blue{
\subsection{\blue{Consistency Properties}}
Let us now study the consistency properties of proposed criterion. Following~\cite{Mooij16}, a plausible criterion of causality for ANMs has to rely on a consistent dependence criterion, and the combination of the ANM and the dependence estimator should be consistent too. \\
Let us show first that the norm of the sensitivity map acts as a consistent dependence measure. Note that in \cite{Hoyer08} originally proposed to use the $p$-value of HSIC to define the causality criterion which leads to a consistent procedure, and in~\cite{Mooij16} it is shown that the same happens for the HSIC itself. Therefore, we have to ensure here that the norm of the sensitivity maps acts as a consistent dependence measure. It is customary to consider that if an estimator $\rho(X,Y)$ fulfills that ``$\rho^{\ast}$(X,Y)$=0$ iff $X$ and $Y$ are statistically independent'', %Property 4 
    then it is considered a dependence measure. Let us first give an empirical demonstration that the norm of the sensitivity maps of HSIC is a consistent estimator: under the hypothesis of independence (Fig.~\ref{fig:1}[a-b]) both HSIC and $\|$SHSIC$\|$ converge to zero with $N$; while in an example of increasing dependence (Fig.~\ref{fig:1}[c])  convergence is ensured too.\\ 
    In theoretical terms, to ensure consistency of the proposed criterion, we must demonstrate that for each of the four terms in 
    $$\widehat C_s: = C_b-C_f = (s_b^y + s_b^{r})-(s_f^x + s_f^{r}).$$ 
    We will show it for the forward term $C_f$, the same arguments apply for $C_b$:
    $$C_f(x,r):= s_f^x+s_f^r = \|\partial_x \text{HSIC}\|^2+\|\partial_r \text{HSIC}\|^2.$$
    Note that for real numbers, the norm of the sensitivity maps of HSIC should be Lipschitz continuous to ensure convergence, that is, we need to demonstrate $|C_f(x,r)-C_f(x,r')|\leq L \|r-r'\|$. For that, we can show that a multivariate function $f:\Real^D\to\Real$ with bounded partial derivatives, which is our case, is Lipschitz, %is Lipschitz iff there exists a constant $L$ such that the restriction of $f$ to every line parallel to a coordinate axis is Lipschitz with constant $L$, 
    and that the bound is exactly $L=\sqrt{2}\max_i(\sup|\partial_{r_i} \text{HSIC}|)$, $i=1,\ldots,N$. This is an equivalent result to Lemma 16 in~\cite{Mooij16} for the HSIC sensitivity map.\\
The second condition to fulfill is that the norm of the sensitivity map in ANMs should be consistent. This can be demonstrated following the same procedure that the one in Theorem 20 (Appendix A.3) in~\cite{Mooij16} for HSIC, and using the previous consistency lemma for the proposed estimator here.
}

\subsection{\blue{Computational cost and efficient criterion}}

\blue{Note that both HSIC and its sensitivity map give raise to closed-form solutions, just involving simple matrix multiplication and a trace operation.  Both HSIC and its sensitivity scale quadratically with the number of examples $N$ since the involved kernel matrices and the centering matrix are $N\times N$. This makes both HSIC and its sensitivity map unfeasible with moderate to large datasets.  In~\cite{PEREZSUAY2017} we provided with fast versions of both HSIC and its sensitivity map through the use of random Fourier features. The cost of a naive implementation of HSIC is ${\mathcal O}(N^3)$, and its randomized version is ${\mathcal O}(D^3)$, where $D$ is the number of Fourier features chosen to approximate the kernel matrices, which is typically smaller than the number of points, i.e. $D\ll N$. Respectively, the naive implementation of the sensitivity of HSIC is ${\mathcal O}(N^3+N^2)\approx{\mathcal O}(N^3)$ and its randomized version scales as ${\mathcal O}(D^3+D^2N)$ so in the cases of $N\gg D$ the cost reduces to ${\mathcal O}(D^2N)$. The cost of the causal criteria is thus defined by these operations. In addition, we demonstrated the convergence of both the randomized HSIC to HSIC, and their corresponding sensitivity maps, which allows a practical use of the method.}

\section{Results} \label{sec:results}

In this section we show the performance of the proposed methodology in three experimental settings: (1) in a collection of 28 geoscience causal inference problems, (2) in a database of radiative transfer models simulations and machine learning emulators in Sentinel-2 vegetation parameter modeling conforming a set of 182 causal problems with groundtruth, and (3) in assessing the impact of different regression models based on GPs in discovering causality in Net Ecosystem Production variables. All problems are bivariate and a annotated ground truth is available based on expert knowledge or common sense, which allows us to assess performance quantitatively.

We quantify the accuracy in detecting the direction of causation using standard scores like the receiver operating curves (ROC), precision-recall curves (PRC) curves, and the areas under these curves. We compare in all cases the state-of-the-art criterion in \eqref{eq:C} with our criterion in \eqref{eq:Cs}. Methods performance are also studied under different situations of number of available points, presence of different noise sources and distortions, and impact of different GP regression models on the detection accuracy.

\begin{figure*}[t!]  % 1:100,[52:55 71]  foreach
\foreach \i in {1,2,3,4,20,21,42,43,44,45,46,49,50,51,72,73,78,79,80,81,82,83,87,89,90,91,92,93} {
\includegraphics[width=2cm]{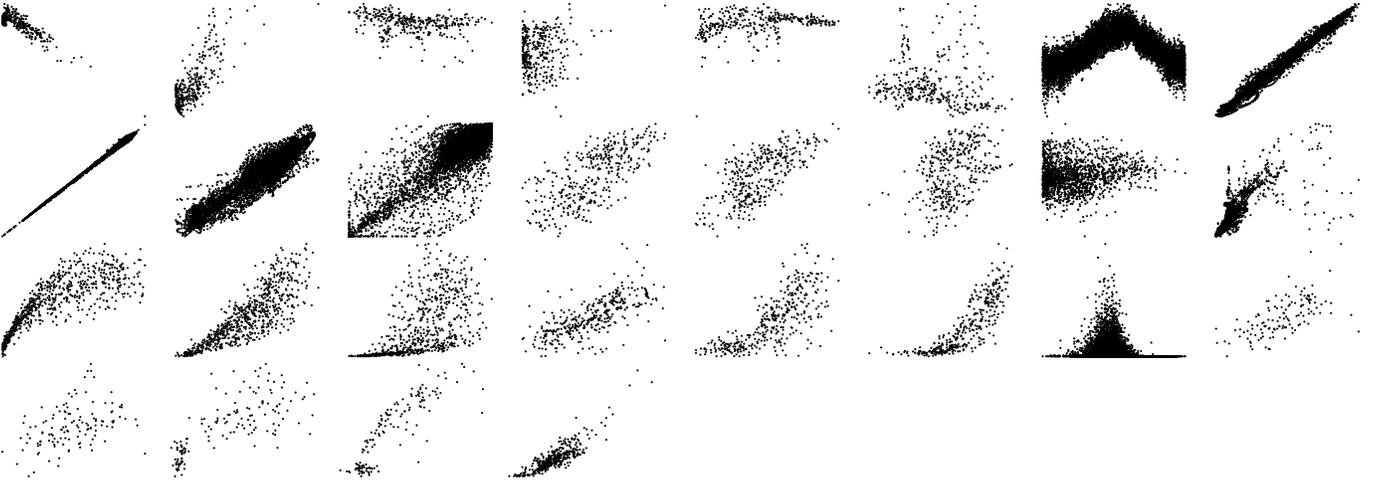}
}
\caption{Scatter plots in the CEP benchmark dataset. See a brief description of all individual problems in Table~\ref{tab:cep} and more details in~\cite{Mooij16jmlr}. \label{fig:subfig}}
\end{figure*}

%%%%%%%%%%%%%%%%%%%%%%%%%%%%%%%%%%%%%%%%%%%%%%%%%%%%%%%%%%%%%%%%%%
\subsection{Experiment 1: Geoscience Cause Effect Pairs}
%%%%%%%%%%%%%%%%%%%%%%%%%%%%%%%%%%%%%%%%%%%%%%%%%%%%%%%%%%%%%%%%%%

\begin{table}[t!]
\begin{center}
\caption{Problems and causal direction for the geoscience problems in the CEP database.\label{tab:cep}}
\setlength{\tabcolsep}{2pt}
\small
\begin{tabular}{|l|l|l|l|}
\hline
id & $x$ & $y$ & Cause \\
\hline
01  &  Altitude  &  Temperature  &  $\rightarrow$ \\
02  &  Altitude  &  Precipitation  &  $\rightarrow$ \\
03  &  Longitude  &  Temperature  &  $\rightarrow$ \\
04  &  Altitude  &  Sunshine hours  &  $\rightarrow$ \\
20  &  Latitude  &  Temperature  &  $\rightarrow$ \\
21  &  Longitude  &  Precipitation  &  $\rightarrow$ \\
42  &  Day of the year  &  Temperature  &  $\rightarrow$ \\
43  &  Temperature at $t$  &  Temperature at $t+1$  &  $\rightarrow$ \\
44  &  Pressure at $t$  &  Pressure at $t+1$  &  $\rightarrow$ \\
45  &  Sea level pressure at $t$  &  Sea level pressure at $t+1$  &  $\rightarrow$ \\
46  &  Relative humidity at $t$  &  Relative humidity at $t+1$  &  $\rightarrow$ \\
49  &  Ozone concentration  &  Temperature  &  $\leftarrow$ \\
50  &  Ozone concentration  &  Temperature  &  $\leftarrow$ \\
51  &  Ozone concentration  &  Temperature  &  $\leftarrow$ \\
72  &  Sunspots  &  Global mean temperature  &  $\rightarrow$ \\
73  &  CO2 emissions  &  Energy use  &  $\leftarrow$ \\
77  &  Temperature  &  Solar radiation  &  $\leftarrow$ \\
78  &  PPFD  &  Net Ecosystem Productivity  &  $\rightarrow$ \\
79  &  NEP  &  Diffuse PPFDdif  &  $\leftarrow$ \\
80  &  NEP  &  Diffuse PPFDdif  &  $\leftarrow$ \\
81  &  Temperature   &  Local CO2 flux, BE-Bra  &  $\rightarrow$ \\
82  &  Temperature   &  Local CO2 flux, DE-Har  &  $\rightarrow$ \\
83  &  Temperature   &  Local CO2 flux, US-PFa  &  $\rightarrow$ \\
87  &  Temperature  &  Total snow  &  $\rightarrow$ \\
89  &  root decomposition  &  root decomposition (grassland)  &  $\leftarrow$ \\
90  &  root decomposition  &  root decomposition (forest)  &  $\leftarrow$ \\
91  &  clay content in soil  &  soil moisture  &  $\rightarrow$ \\
92  &  organic carbon in soil  &  clay cont. in soil (forest)  &  $\leftarrow$ \\
93  &  precipitation  &  runoff  &  $\rightarrow$ \\
94  &  hour of day  &  temperature  &  $\rightarrow$ \\
\hline
\end{tabular}
\end{center}
\end{table}

We used Version 1.0 of the CauseEffectPairs (CEP) collection freely\footnote{\url{https://webdav.tuebingen.mpg.de/cause-effect/}}. The database contains 100 pairs of random variables along with the right direction of causation (ground truth). Data has been collected from various domains of application, such as biology, climate science, health sciences and economics, just to name a few. More information about the dataset and an excellent up-to-date review of observational causal inference methods is available in~\cite{Mooij16jmlr}. We conducted experiments in 28 out of the 100 pairs that contain one-dimensional variables and that are related to geosciences and remote sensing: problems involving carbon and energy fluxes, ecological indicators, vegetation indices, temperature, moisture, heat, etc. We summarize the involved variables in Table~\ref{tab:cep}, and show the scatter plots of the selected pairs $(x, y)$ in Fig.~\ref{fig:subfig}. %Each problem in the dataset has different amount of examples, noise amounts and sources.

\subsubsection{Experimental Setup}

The experimental setting is as follows. Once the two predictive models $f$ and $g$ were developed, we computed the two HSIC terms, as well as their sensitivity maps between the $x$ (or $y$) and the residuals $r_f$ (or $r_b$). The final causal direction score was simply defined as the difference in test statistic between both models, either using $\hat C$ or the proposed $\hat C_s$. Note that this is a particular form of `ranked-decision' setting that needs to account for the bias introduced by pairs coming from the same problem, i.e. it is customary to down-weight the precision for every decision threshold in the curves (e.g. four related problems receive 0.25 weights in the decision)\footnote{The MATLAB function \url{perfcurve} can produce such (weighted) ROC and PRC curves and the estimated weighted AUC.}. This is the case, for example of problems 81, 82, 83 and 87 that receive 0.25 weights.

\subsubsection{Accuracy and robustness of the detection}

\begin{figure}[h!]
\setlength{\tabcolsep}{1pt}
\begin{center}
%\begin{tabular}{cccc}
\includegraphics[width=8.4cm]{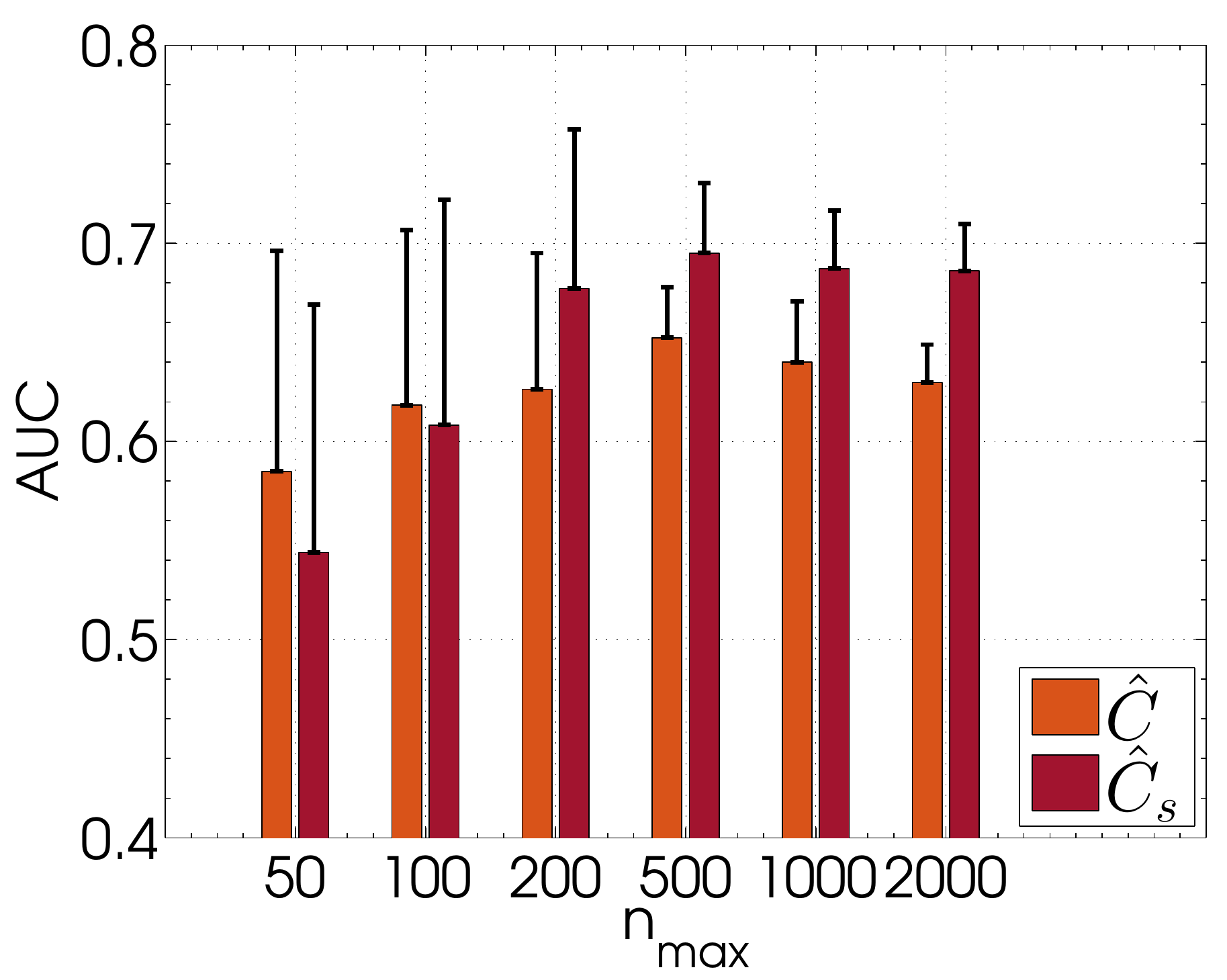} % & 
%\includegraphics[width=4.2cm]{images/28problems/PRbars_IGARSS17.pdf}  & 
%\end{tabular}
\end{center}
%\vspace{-0.7cm}
\caption{AUCs in the CEP causality problems dataset for different amounts of training data per problem. \label{fig:sensrocbars}}
\end{figure}

\begin{figure}[h!] 
\setlength{\tabcolsep}{1pt}
\begin{center}
%\begin{tabular}{cc}
\includegraphics[width=7.4cm]{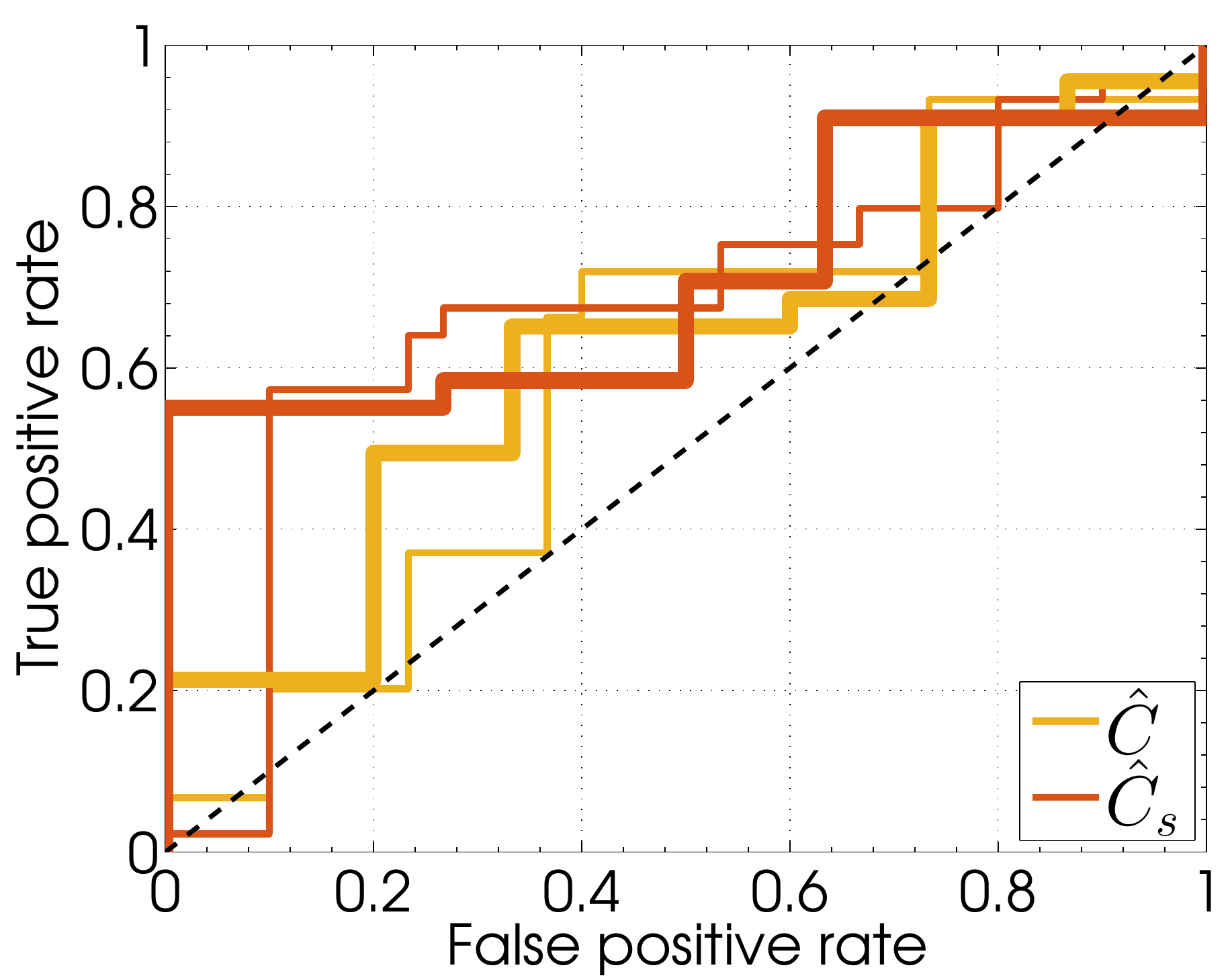} \\%&
\includegraphics[width=7.4cm]{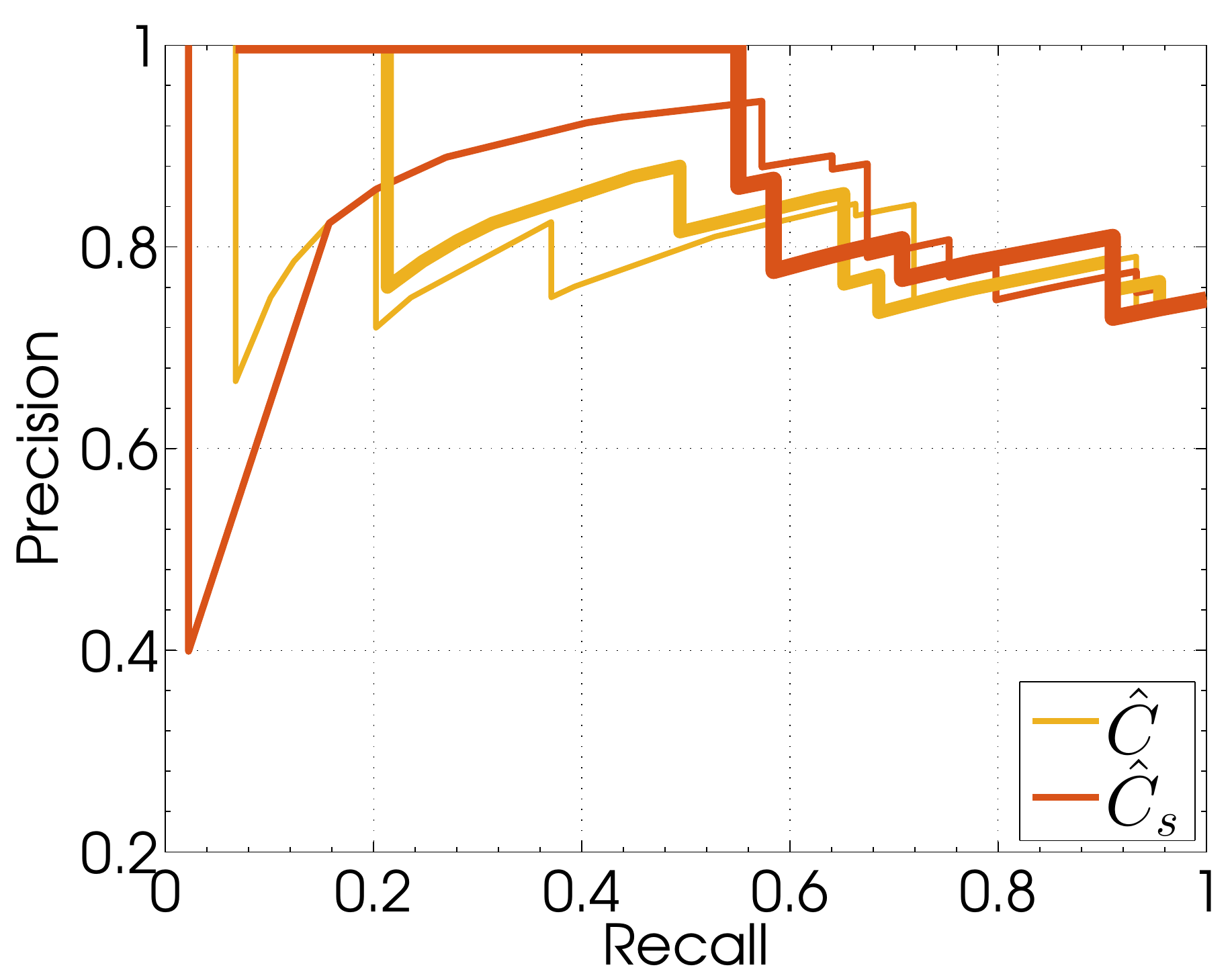} 
%\end{tabular}
\end{center}
\caption{\blue{Receiver operating curve (ROC) (top) and Precision-Recall (PR) (bottom) curves for the causality problems in the Geoscience Cause Effect Pairs (experiment section 1) using $n=200$ (thin lines) and $n=2000$ (thick lines). Higher area under the curves are obtained with the proposed criterion $C_s$ and for both situations.} \label{fig:roc}}
\end{figure}

We run the experiments with different numbers of (randomly selected) points $n$ from both variables. This situation impacts regression models performance, both in terms of the regression accuracy and the dependence estimation. We evaluate $\hat C$ %\red{$C_p$}, 
and $\hat C_s$ criteria by limiting the maximum number of training samples in each problem, $n_{max}=\{50,100,200,500,2000\}$. Results were averaged over $10$ realizations. Figure~\ref{fig:sensrocbars} shows the AUC under the curve as a function of $n_{max}$. The proposed sensitivity-based criterion consistently performed better than the standard approach using HSIC alone. \blue{Lower values of AUC are obtained by our proposal only when the number of samples is relatively low (50 and 100). This particular behavior is plausible because our criterion is defined through an empirical estimator (the sensitivity) and when the number of samples is moderately low it can give rise to underestimated values.} Looking more in detail at the ROC in Fig.~\ref{fig:roc}, we note that the proposed $\hat C_s$ yields the best recognition curves, both ROC and the PRC. Note that this happens for all decision rates, especially in low false positive rates regimes, and for all number of training points. 

%%%%%%%%%%%%%%%%%%%%%%%%%%%%%%%%%%%%%%%%%%%%%%%%%%%%%%%%%%%%%%%%%%
\subsection{Experiment 2: Causation in RTM assessment}
%%%%%%%%%%%%%%%%%%%%%%%%%%%%%%%%%%%%%%%%%%%%%%%%%%%%%%%%%%%%%%%%%%

Using input-output data pairs generated by radiative transfer models (RTMs) allow us to assess performance of observational causality algorithms: it is obvious that the forward RTM simulation gives the right direction of causation: state vectors (parameters) cause radiances and not the other way around. The strength of the causation is not considered here.  In this second experiment, we deal with data generated by the standard PROSAIL RTM\footnote{\url{http://teledetection.ipgp.jussieu.fr/prosail/}}. PROSAIL is the combination of the PROSPECT leaf optical properties model and the SAIL canopy bidirectional reflectance model. PROSAIL has been used to develop new methods for retrieval of vegetation biophysical properties. Essentially, PROSAIL links the spectral variation of canopy reflectance, which is mainly related to leaf biochemical contents, with its directional variation, which is primarily related to canopy architecture and soil/vegetation contrast. This link is key to simultaneous estimation of canopy biophysical/structural variables for applications in agriculture, plant physiology, and ecology at different scales. PROSAIL has become one of the most popular radiative transfer tool due to its ease of use, robustness, and consistent validation by lab/field/space experiments over the years. 

\begin{table}[t!]
\small
\renewcommand{\tabcolsep}{1pt}
\caption{Configuration parameters of the simulated data.\label{tab:prosail}}
%\vspace{-0.6cm}
\begin{center}
\begin{tabular}{|l|l|l|l|}
\hline
\hline
Parameter &	Sampling & Min & Max \\
\hline
\hline
\multicolumn{4}{|l|}{RTM model: Prospect 4} \\
\hline
Leaf Structural Parameter            				&	Fixed                      	& 	1.50	& 1.50 \\
C$_{ab}$, chlorophyll a+b [$\mu$g/cm$^2$]  & 	${\mathcal U}(14,49)$      	& 	0.067	& 79.97 \\
C$_w$, equivalent water thickness [mg/cm$^2$] 			&	${\mathcal U}(10,31)$ 	&	2	& 50 \\
C$_m$, dry matter [mg/cm$^2$]         			& 	${\mathcal U}(5.9,19)$ 	&	1.0 	& 3.0 \\
\hline
\multicolumn{4}{|l|}{RTM model: 4SAIL} \\
\hline
Diffuse/direct light    &	Fixed         &                                      	10 &	10\\
Soil Coefficient       & 	Fixed         &                                      	0 &	0\\
Hot spot               &	Fixed         &                                      	0.01 &	0.01\\
Observer zenit angle     &	Fixed         &                                      	0 &	0\\
LAI, Leaf Area Index    &	${\mathcal U}(1.2,4.3)$ &   0.01 &	6.99\\
LAD, Leaf Angle Distribution  &	${\mathcal U}(28,51)$ &   20.04 &	69.93\\
SZA, Solar Zenit Angle        &	${\mathcal U}(8.5,31)$  &   0.082 &	49.96\\
PSI, Azimut Angle             &	${\mathcal U}(30,100)$ &    0.099 &	179.83\\
\hline
\hline
\end{tabular}
\end{center}
\end{table}

\subsubsection{Experimental Setup}

We used PROSAIL to generate $n=1000$ pairs of Sentinel-2 spectral (13 spectral channels) by varying 7 parameters in reasonable ranges: Total Leaf Area Index (LAI), Leaf angle distribution (LAD), Solar Zenit Angle (SZA), Azimut Angle (PSI), chlorophyll a+b content $C_{ab}$ [$\mu$g/cm$^2$], equivalent water thickness $C_w$ [g/cm$^2$] and dry matter content, $C_m$ [g/cm$^2$]. Several parameters were kept fixed in the simulations. See Table~\ref{tab:prosail} for the configuration details used in our PROSAIL simulations. Building the database assumes that  every individual parameter impacts (causes) a particular spectral channel, and that spectral channels cannot cause the parameters. This returns a simulated dataset with $2\times 13 \times 7=182$ causal problems with ground truth. 

\subsubsection{Accuracy and robustness of the detection}

We run the different criteria and obtained the corresponding ROCs and AUCs. We did observe very high accuracy levels for all criteria. %, essentially because of the ease of the link and the absence of noise. 
We then assessed robustness of the methods to different additive white Gaussian noise levels, varying the SNR in the range [0,40] dB. Figure~\ref{fig:prosail_ROC_PR} shows the obtained results for all causal criteria and accuracy measures as a function of SNR. Our proposed criterion $\widehat{C}_s$ shows better performance than $\widehat{C}$ for all SNR levels, with an average improvement of +5\% in AUC. Both criteria degrade in scenarios dominated by noise (SNR$<$10dB), where neither the functions nor independence can be estimated correctly. %, suggesting that the sensitivity measure may lead to biased estimates in such settings and with low number of samples.

\iffalse  % RESULTS FOR LANDSAT!!!
\begin{figure}[t!]
\setlength{\tabcolsep}{1pt}
\begin{center}
\begin{tabular}{cccc}
\includegraphics[width=4.6cm]{images/prosail/ROC_prosail_noise.pdf}  & 
\includegraphics[width=4.6cm]{images/prosail/PR_prosail_noise.pdf}  & 
\end{tabular}
\end{center}
\caption{ROC (left) and Precision-Recall (right) adding noise. \label{fig:prosail_ROC_PR}}
\end{figure}
\fi

\begin{figure}[h!] 
\setlength{\tabcolsep}{1pt}
\begin{center}
%\begin{tabular}{cc}
\includegraphics[width=7.4cm]{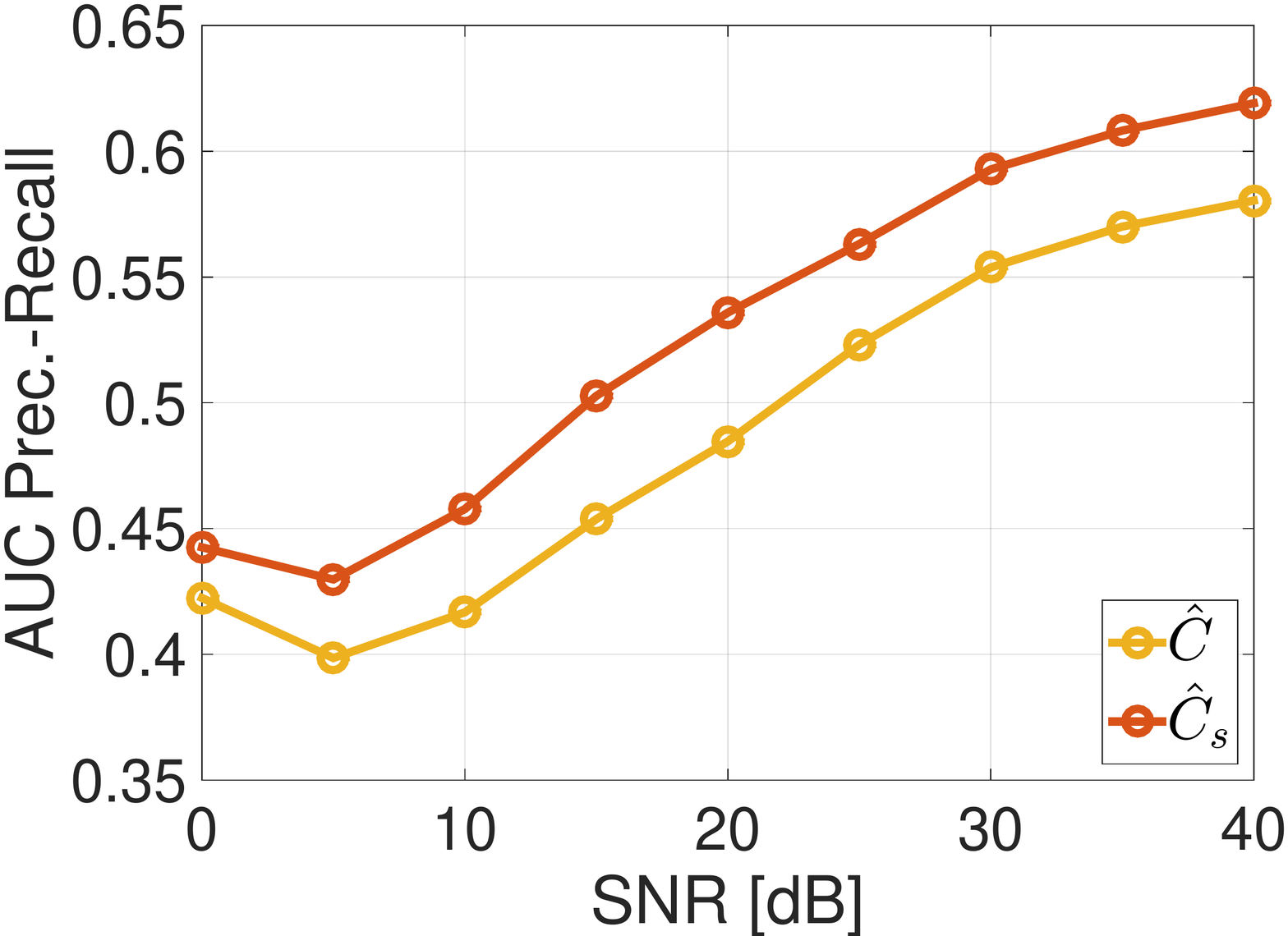} \\% &
\includegraphics[width=7.4cm]{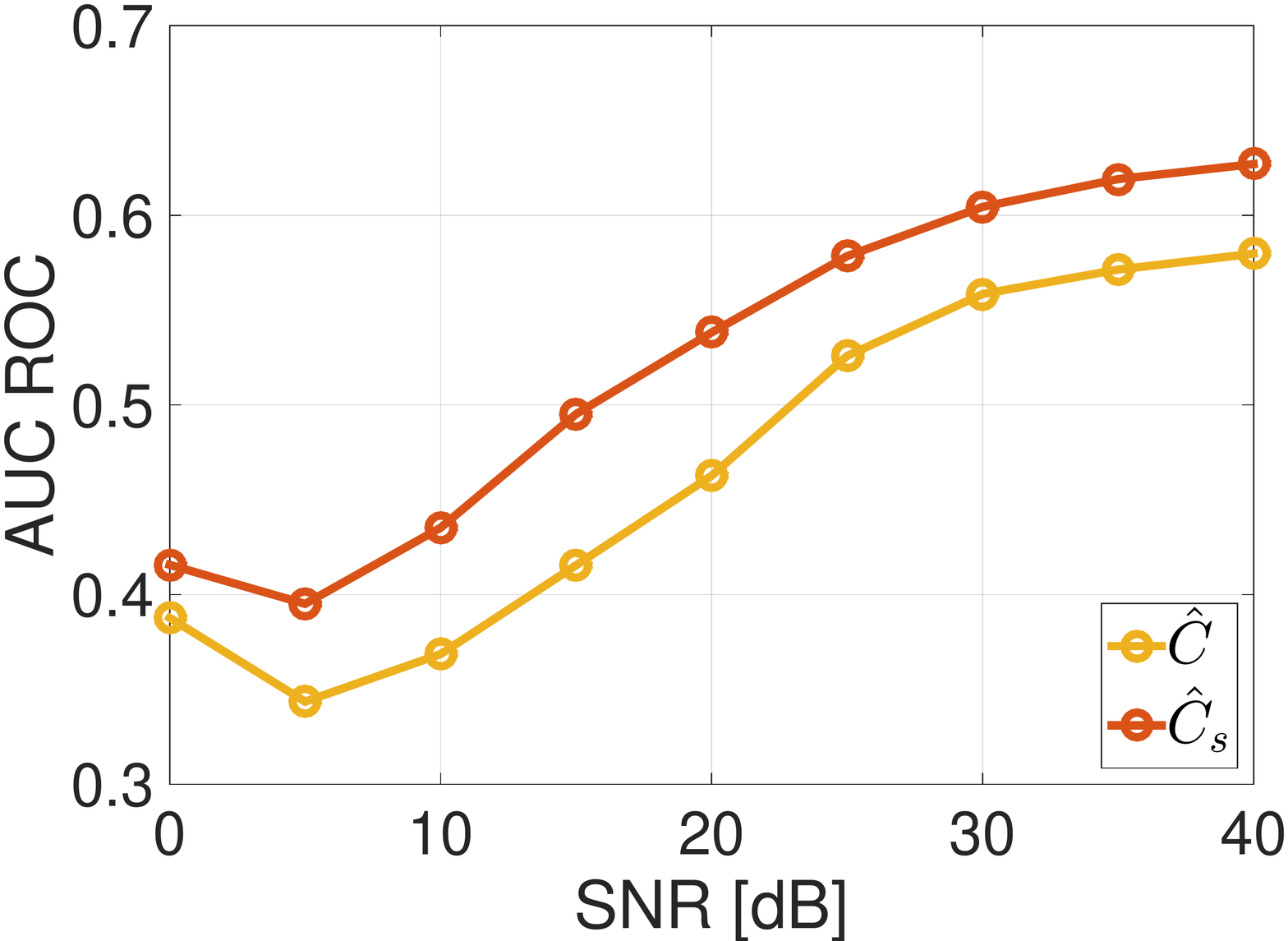} 
%\end{tabular}
\end{center}
\caption{\blue{Area under the curve (AUC) of Precision-Recall (top) and AUC of the ROC curves (bottom) both against the amount of noise included in the experiments. Higher detection power is obtained with the proposed $C_s$ for all noise levels and measures.} \label{fig:prosail_ROC_PR}}
\end{figure}

\iffalse
Figure~\ref{fig:criteriamaps} shows all pairwise values of the derived criteria $C$ and $C_s$. In general:
\begin{itemize}
\item C$_s$ yields sharper identification of relations
\item Canopy structural variable LAI pops up from NIR on
\item Causal link LAD$\to$[NIR, 1500nm]
\item Chlorophyll content (Cab) drives reflectance in the visible
\item Leaf water content (Cw) governs leaf reflectance $>1200$ nm
\item Dry matter content (Cm) are key all over %, key across all spectral range
\end{itemize}

\begin{figure}[t!]
\begin{center}
\small
\setlength{\tabcolsep}{1pt}
\begin{tabular}{cc}
{$\hat C$ (AUC=0.74)} & {$\hat C_s$ (AUC=0.76)} \\[-2mm]
\includegraphics[width=4.4cm]{./images/prosail/HSICsens.pdf} & 
\includegraphics[width=4.4cm]{./images/prosail/Cssens.pdf} \\
\end{tabular}
\end{center}
\caption{cccc\ref{fig:criteriamaps}}
\end{figure}
\fi

%%%%%%%%%%%%%%%%%%%%%%%%%%%%%%%%%%%%%%%%%%%%%%%%%%%%%%%%%%%%%%%%%%
\subsection{Experiment 3: Causation in RTM Emulation}
%%%%%%%%%%%%%%%%%%%%%%%%%%%%%%%%%%%%%%%%%%%%%%%%%%%%%%%%%%%%%%%%%%

As observed before, noise plays a fundamental role in causal discovery. In this experiment we studied the impact of other types of distortions in remote sensing data. In particular, we aim to assess the non-linearities introduced when approximating a physical model via machine learning. This form of surrogate modelling is known in the literature as {\em emulation}, and has captured the attention in recent years because it allows to replace RTMs with more efficient statistical algorithms~\cite{CampsValls16grsm,Verrelst12emula}. Emulators, however are just function approximators and the simulated radiances are subject to complicated distortions. We evaluate the identifiability of the causal links in such cases.

\begin{figure*}[t!]
\setlength{\tabcolsep}{1pt}
\begin{center}
\begin{tabular}{cccc}
\includegraphics[width=6.2cm]{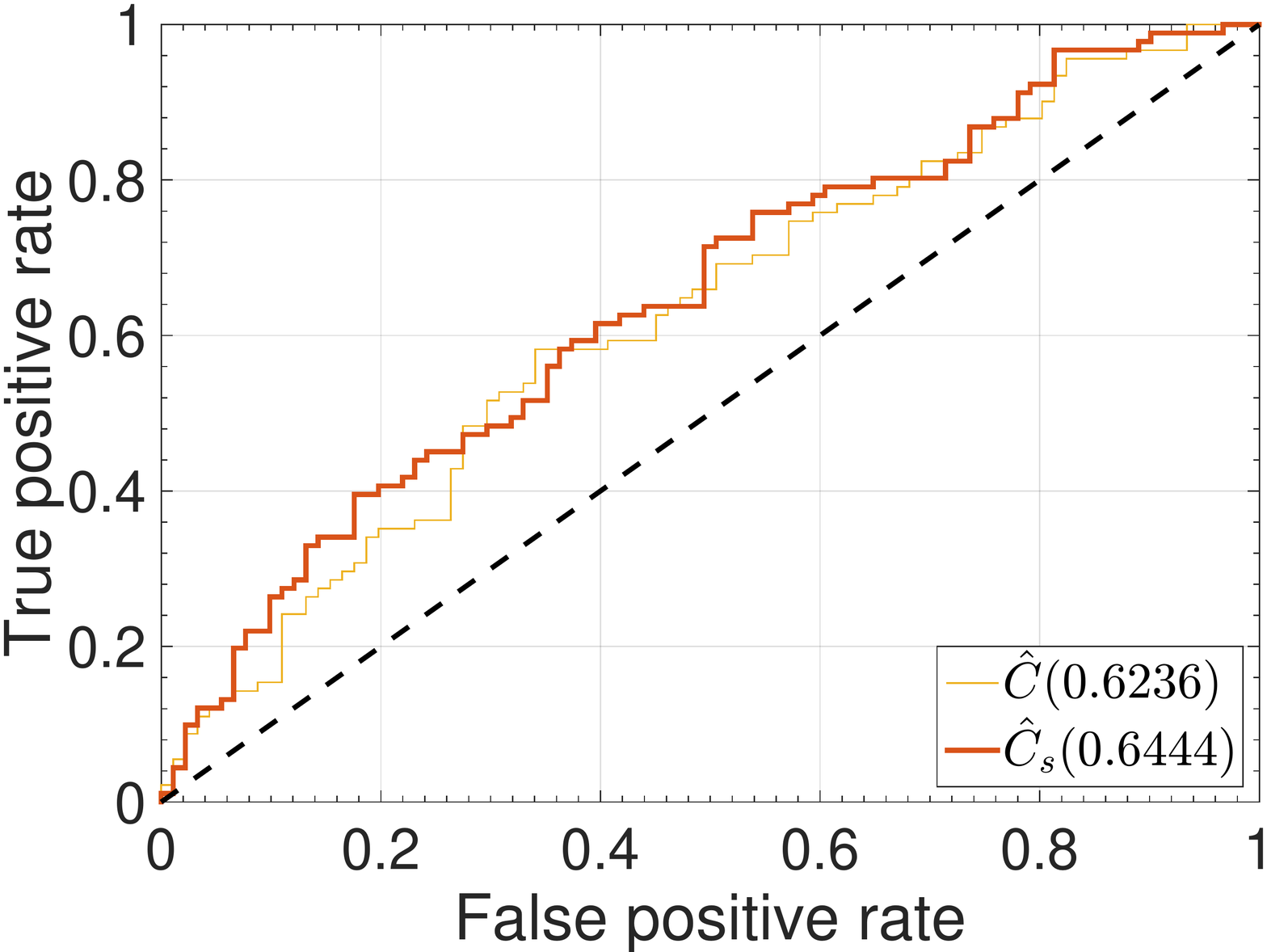}  & 
\includegraphics[width=6.2cm]{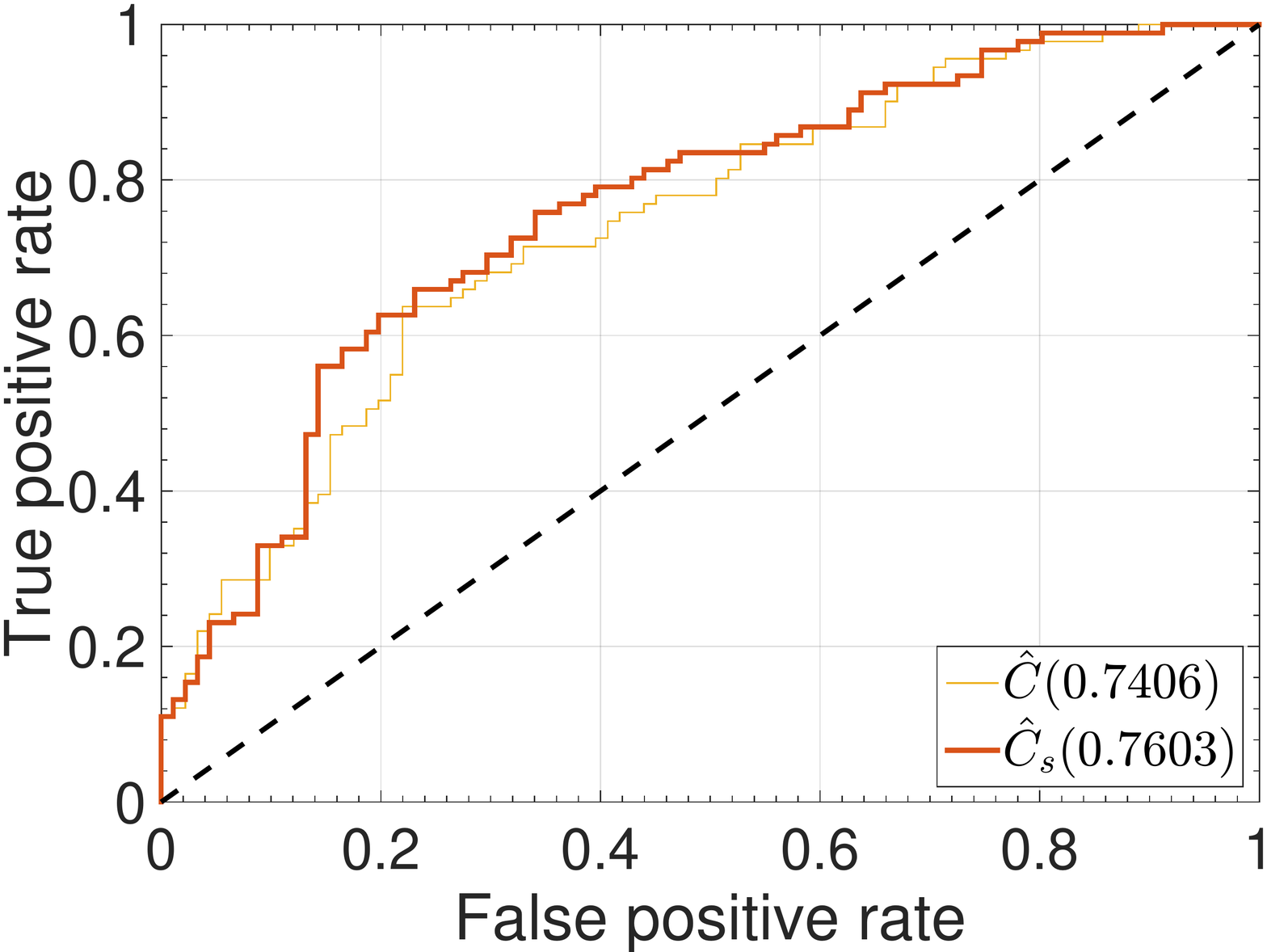}  & \includegraphics[width=6.2cm]{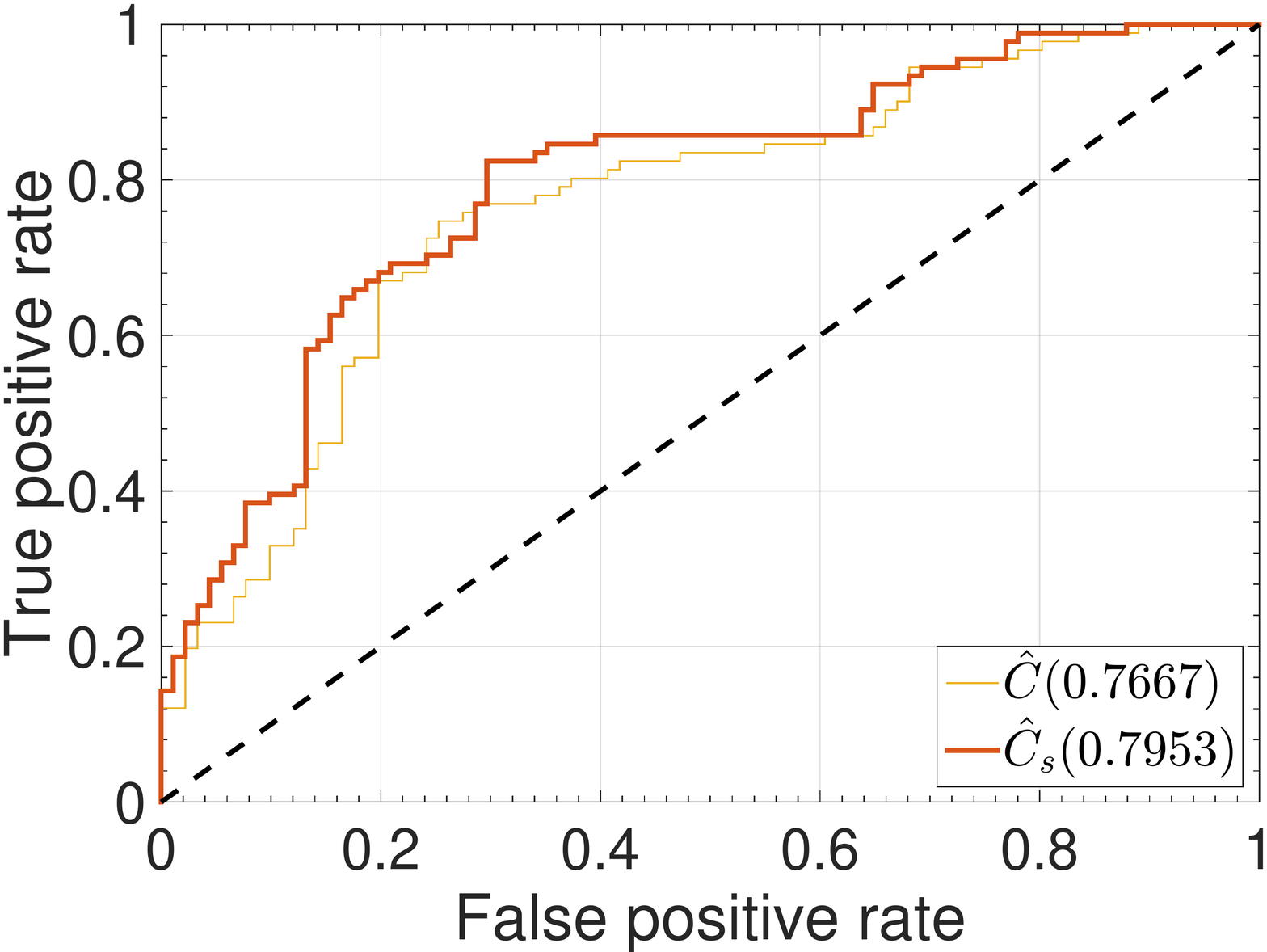}\\
\includegraphics[width=6.2cm]{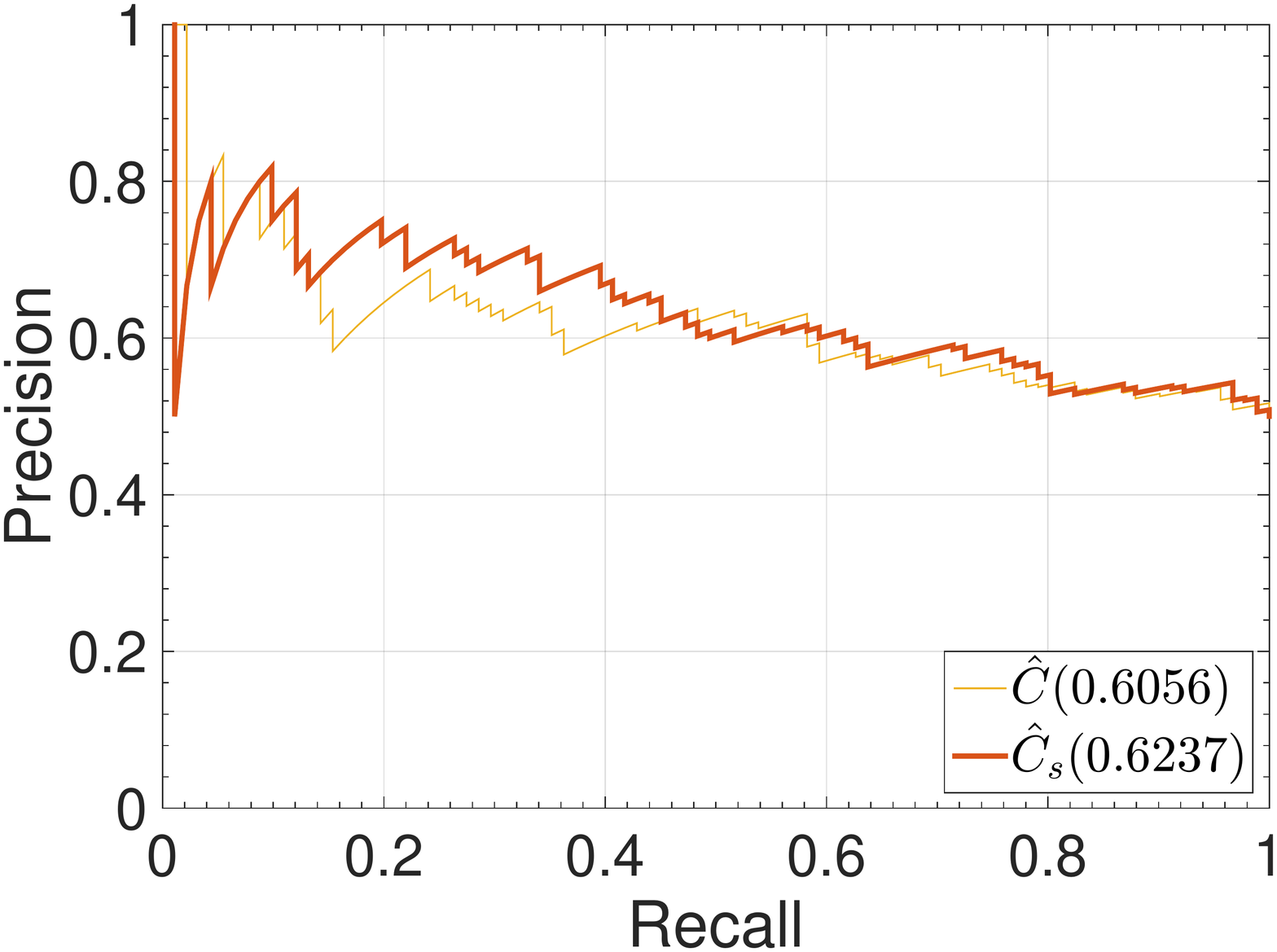}  & 
\includegraphics[width=6.2cm]{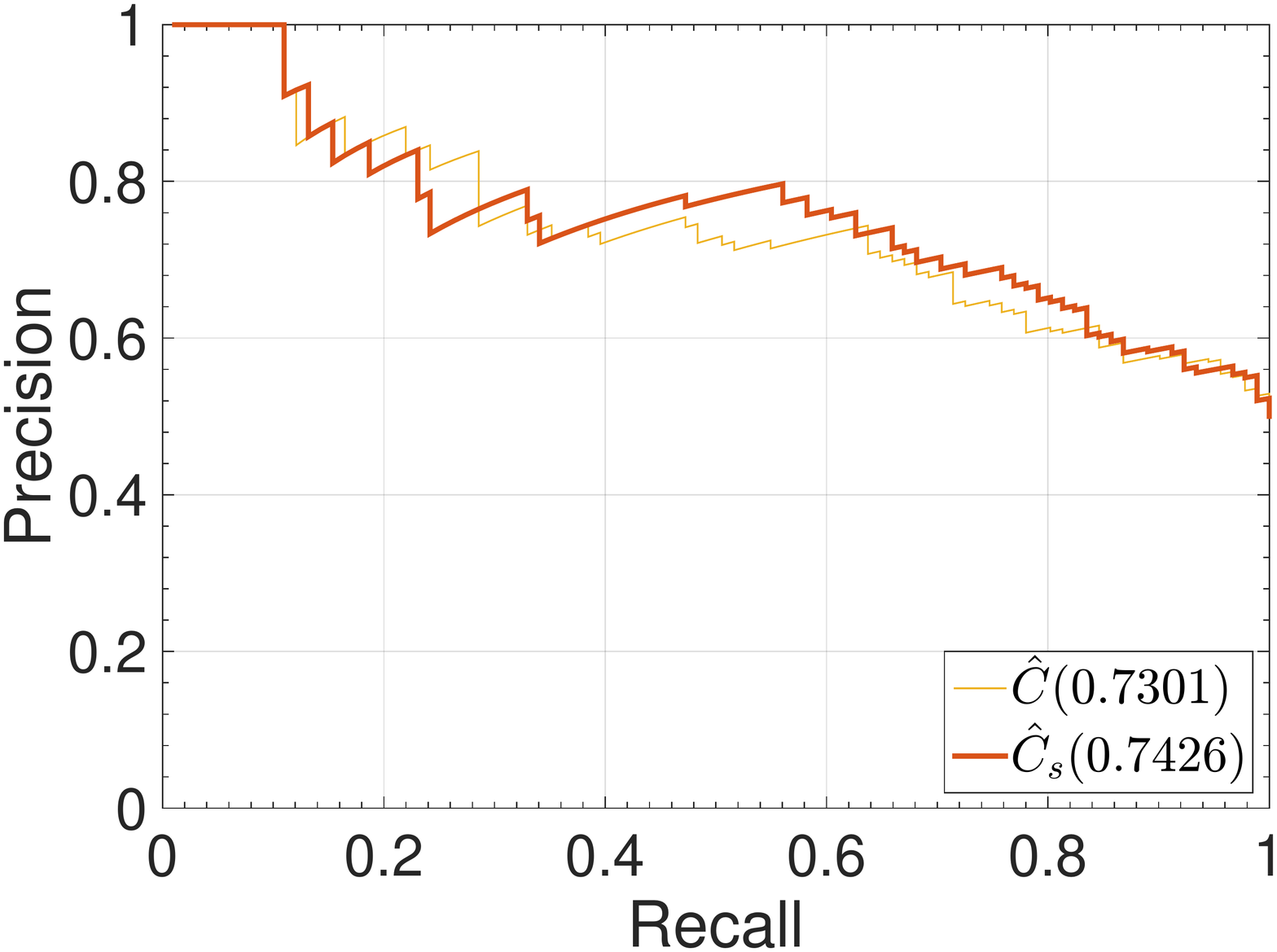}  & \includegraphics[width=6.2cm]{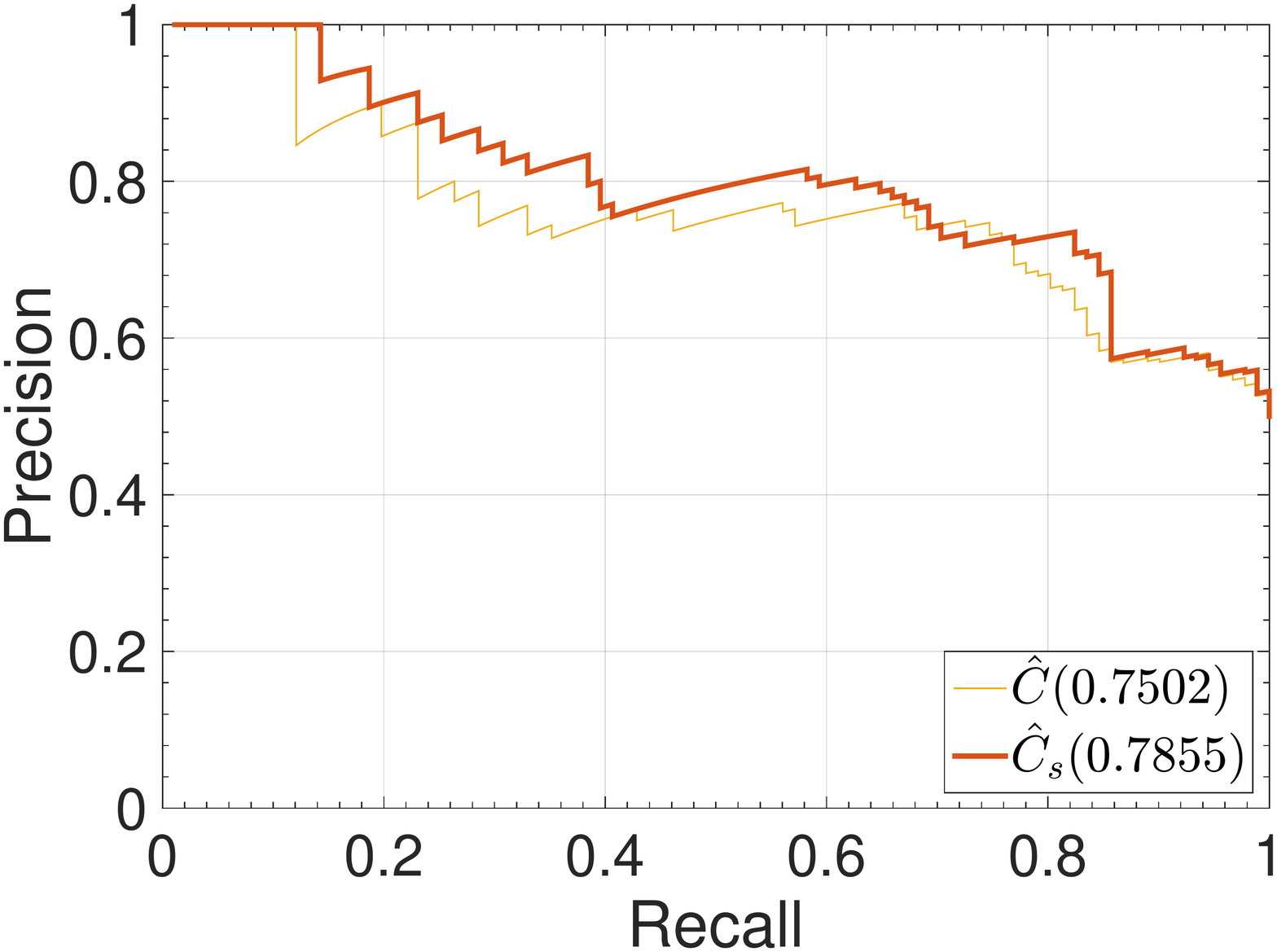}\\
\end{tabular}
\end{center}
\caption{ROC (top row) and Precision-Recall (bottom row) curves for the emulation database of 182 causality problems.   \label{fig:roc2}}
\end{figure*}

Here we trained a neural network using the $n=1000$ points generated by PROSAIL in the previous experiment to build an emulator. Performance showed less than 5\% of normalized RMSE in all bands. Once trained, the emulator was run to generate $n=10^6$ samples. The same amount of 182 cause-effect bivariate problems as before was generated.  
We run the different criteria training the regression models with different amounts of data points, $n=\{10000,20000,30000\}$, and the standard AUC and PR criteria are shown in Fig.~\ref{fig:roc2}. Results show that (1) all criteria improve performance with an increasing amount of data, and (2) our criterion $\widehat{C}_s$ outperforms the state-of-the-art $\widehat{C}$ in all cases (by around +2-4\%). % $2\times 13 \times 7=182$ labeled problems of cause-effect pairs by simple randomization. It is physics understanding (and common sense) that state vectors are the causes and the radiance observations are the effects: the standard forward model in RTM simulation gives us the causal direction. 

%\begin{figure}[t!]
%\setlength{\tabcolsep}{1pt}
%\begin{center}
%\begin{tabular}{cccc}
%\includegraphics[width=4.2cm]{images/28problems/ROC_EMULATOR_IGARSS17.pdf}  & 
%\includegraphics[width=4.2cm]{images/28problems/PR_EMULATOR_IGARSS17.pdf}  & 
%\end{tabular}
%\end{center}
%\caption{ROC (left) and Precision-Recall (right) curves for the PROSAIL database of 182 causality problems %\red{Ummmm; que passa amb la ROC?} \label{fig:roc2}}
%\end{figure}

%%%%%%%%%%%%%%%%%%%%%%%%%%%%%%%%%%%%%%%%%%%%%%%%%%%%%%%%%%%%%%%%%%
\subsection{Experiment 4: Impact of the regression model}
%%%%%%%%%%%%%%%%%%%%%%%%%%%%%%%%%%%%%%%%%%%%%%%%%%%%%%%%%%%%%%%%%%

%\red{REPE???? First, we must note that we explored different regression algorithms in this work to derive both the forward and backward regression models, $f$ and $g$. We explored mainly Gaussian Process Regression (GPR)~\cite{Rasmussen06,CampsValls16grsm} as suggested in~\cite{Mooij16}, as well as some improved GP models, such as the warped GP regression (WGPR)~\cite{Munoz15igarss} and variational heteroscedastic GP regression (VHGPR)~\cite{CampsVallsGRSL2013}. In addition, we used standard random forests (RFs)~\cite{Breiman01}, which in some cases provided more accurate and faster results. See our preliminary results with GPs and RFs in  in~\cite{PerezSuay17igarss,Perez17shsic}. In all cases, we used RBF kernels, and the lengthscales were fixed to the average Euclidean distance among all available data points.}

In this last experiment, we are concerned about the use of different regression algorithms that better account for noise and non-linearities~\cite{CampsValls16grsm}. In particular, we will compare the use of the standard (homoscedastic) GP regression model (GP)~\cite{Rasmussen06} (cf. Section~\ref{sec:causal}), with the heteroscedastic GP model (VHGP) introduced in~\cite{Lazaro11,CampsValls16grsm} (which accounts for signal-to-noise relations), and a warped GP model (WGP) introduced in~\cite{Lazaro12warp,Munoz15igarss} (which further transforms model's output to look more like a Gaussian process). 

We exemplify these different approaches in a relevant geoscience problem. %The last few hundred years, human activities have precipitated an environmental crisis on Earth, commonly described as `global climate change.' Since the discovery of fossil carbon as a convenient form of energy, the residues of past photosynthetic carbon assimilation have been combusted to CO$_2$ and returned to the Earth's atmosphere. 
Terrestrial ecosystems absorb approximately 120 Gt of carbon annually from the atmosphere, about half is returned as plant respiration and the remaining 60 Gt yr$^{-1}$ represent the Net Primary Production (NPP). Out of this, about 50 Gt yr$^{-1}$ are returned to the atmosphere as soil/litter respiration or decomposition processes, while about 10 Gt yr$^{-1}$ results in the Net Ecosystem Production (NEP). The problem here deals with estimating the causal relation between the photosynthetic photon flux density (PPFD), which is a measure of light intensity\footnote{The total PPFD was measured here as the number of photons falling on a one square meter area per second, while NEP was calculated by photosynthetic uptake minus the release by respiration, which is known to be driven by either the total, diffuse or direct PPFD.}, and the NEP, which results from the potential of ecosystems to sequestrate atmospheric carbon. Discovering such relations may be helpful to better understand the carbon fluxes and to establish sinks and sources of carbon. We use here three data sets taken at a flux tower at site DE-Hai involving PPFD(total), PPFD(diffuse), PPFD(direct) drivers and the NEP consequence variable~\cite{Moffat10}.

Results for all three scenarios are shown in Table~\ref{tab:causal2}. %In this case, we only run $C_p$ for the sake of simplicity and because all criteria showed very similar results. Introducing better models generally lead to sharper detections: much lower $p$-values for the forward direction, $p_f$, and similar $p$-values of the backward direction, $p_b$.  
We show the values of HSIC in forward and backward directions, as well as the criteria obtained by all regression models. The heteroscedastic GP accounts for the signal-to-noise relations in a more sensible way, so the dependency estimate becomes slightly more reliable. Nevertheless performance of WGP excels in these particular problems, probably because of the better estimation of conditionals in higher density regions (see Fig. 1 in~\cite{Snelson04WGP}). Future work will involve testing these models in a wider range of applications.

\begin{table}[t!]
\begin{center}
\small
\setlength{\tabcolsep}{1pt}
\caption{Results in the `PPFD causes NEP' causal problem. \label{tab:causal2}}
\begin{tabular}{|l||c|c||c|c||l|}
\hline
\hline
Method & HSIC$_f$ & HSIC$_b$ & $C$  & $C_s$ & Conclusion\\
%Method & $p_f$  &  $p_b$ & Conclusion\\
\hline
%GPR    & $3.86\times 10^{-61}$  & $1.57\times 10^{-119}$  & PPFD(tot)$\to$ NEP \\
%WGPR  & $2.12\times 10^{-50}$  & $3.33\times 10^{-115}$  & PPFD(tot)$\to$ NEP \\
%VHGPR  & $6.11\times 10^{-60}$  & $2.50\times 10^{-109}$  & PPFD(tot)$\to$ NEP \\
%\hline
%\hline
%GPR    & $1.59\times 10^{-11}$  & $1.24\times 10^{-79}$  & PPFD(diff)$\to$ NEP \\
%WGPR  & $1.17\times 10^{-11}$  & $9.40\times 10^{-77}$  & PPFD(diff)$\to$ NEP \\
%VHGPR  & $2.44\times 10^{-12}$  & $9.16\times 10^{-75}$  & PPFD(diff)$\to$ NEP \\
%\hline
%\hline
%GPR    & $2.05\times 10^{-8}$  & $1.56\times 10^{-112}$  & PPFD(dir)$\to$ NEP \\
%WGPR  & $1.20\times 10^{-15}$  & $3.67\times 10^{-110}$  & PPFD(dir)$\to$ NEP \\
%VHGPR  & $3.44\times 10^{-17}$  & $1.01\times 10^{-115}$  & PPFD(dir)$\to$ NEP \\
%\hline
%\hline
%RF    &   5.7277  &  6.8043  &   1.0765 &  -3.9084 & PPFD(tot)$\to$ NEP \\
GP   &   6.7525  & 10.5255  &   3.7729 &   -1.9204 & PPFD(tot)$\to$ NEP \\
VHGP &   6.8220  & 11.7661  &   4.9441 &  -1.8758 & PPFD(tot)$\to$ NEP \\
WGP  &   6.8670  & 12.0412  &   5.1742 &   -1.6784 & PPFD(tot)$\to$ NEP \\
\hline
\hline
%RF    &   7.5236  &  1.4399  &  -6.0837 &   1.7653 & NEP $\leftarrow$ PPFD(diff)\\
GP   &   8.0982  &  2.1020  &  -5.9961 &    0.8917 & NEP $\leftarrow$ PPFD(diff)\\
VHGP &   8.1556  &  2.1865  &  -5.9691 & 0.8674 & NEP $\leftarrow$ PPFD(diff)\\
WGP  &   8.1707  &  2.0475  &  -6.1232 & 0.7270 & NEP $\leftarrow$ PPFD(diff)\\
\hline
\hline
%RF    &   8.5180  &  1.5617  &  -6.9563 &  0.8182 & PPFD(dir)$\to$ NEP\\
GP   &  11.4727  &  1.5806  &  -9.8920 & -0.7110 & PPFD(dir)$\to$ NEP\\
VHGP &  13.0462  &  1.6848  & -11.3614 & -0.6062 & PPFD(dir)$\to$ NEP\\
WGP  &  13.0061  &  1.6028  & -11.4033 & -0.8046 & PPFD(dir)$\to$ NEP\\
\hline
\hline
\end{tabular}
\end{center}
\end{table}

\section{Conclusions} \label{sec:conclusions}

This paper introduced for the first time the issue of observation-based causal inference in bivariate instantaneous problems in remote sensing and geosciences. Approaching this kind of problems requires taking some (strong) assumptions, so results must be taken with extreme caution. Nevertheless, the obtained results confirm that in general causal detection accuracy is well above chance, and opens the field to further experimentation. 

To tackle this challenging problem, we used a simple method based on regression and dependence estimation, and proposed a new criterion based on the sensitivity (derivative) of the dependence estimator instead of the dependence itself. This allows us to better capture the asymmetry of the forward and inverse densities with regard to the causal mechanism. 

State-of-the-art accuracy was obtained in a wide range of situations with known ground truth. We illustrated performance in a collection of 28 geoscience causal inference problems, in a large database of PROSAIL simulations and emulators in vegetation parameter modeling, %in the problem of identifying the atmospheric inversion layer from infrared sounding data, 
and in a carbon cycle problem. %\red{The datasets are available for experimentation by interested readers.}
We evaluated the impact of using different regression models based on Gaussian Processes; as well as assessed identifiability in the presence of different noise sources and distortions. Models performance was evaluated in global terms by measuring the right direction of causation using standard metrics derived from detection curves. 

We would like to finally note that the methodologies proposed here were originally introduced for general-purpose applications. We have nevertheless shown its applicability in remote sensing and geosciences. \blue{We relied on a general assumption of structural models in general and ANM in particular. If the assumptions are not fulfilled, the method should not perform well. This may happen in some cases, such as in cases of post-nonlinear effects. Actually, we showed this case experimentally where several regression models were used. The WGP generalizes does not assume an additive noise model in general and results actually confirm that, by replacing the regression model, one can achieve more robust results in cases where assumptions are not met.} 

\blue{The proposed scheme for bivariate causal inference can actually include many independence criteria, such as e.g.   differential  entropies, empirical Bayes  scores or minimum message length
scores. We however restricted ourselves to the (standard use of) HSIC, and} included a novel causal criterion (the sensitivity map) that can be computed analytically from HSIC. Other more sophisticated criteria could be actually derived from the combination of the sensitivity map and the HSIC values and $p$-values, which will be matter of future research.

% Adrian
\begin{IEEEbiography}[{\includegraphics[width=1in,height=1.25in,keepaspectratio]{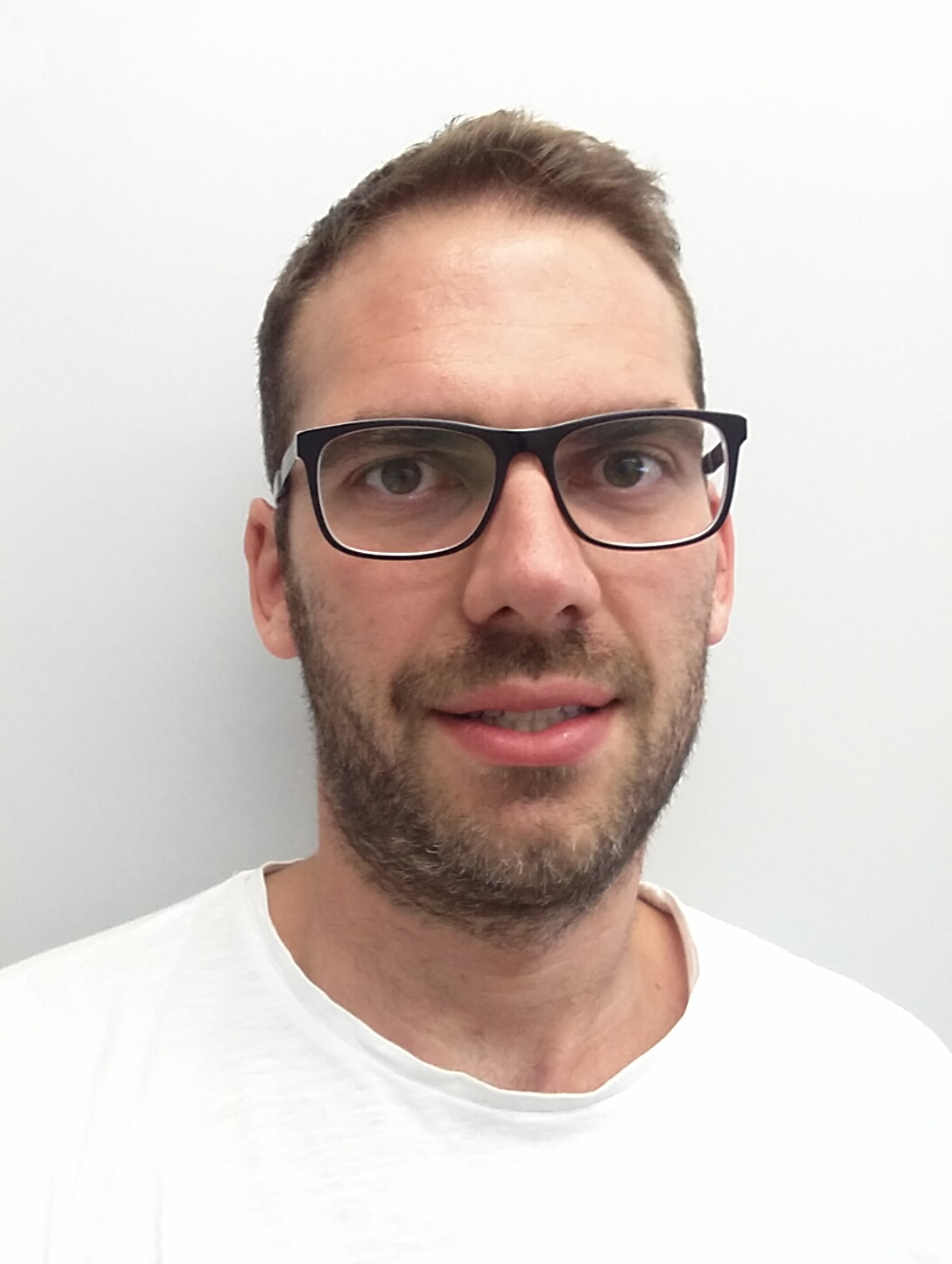}}]{Adri\'{a}n P\'{e}rez-Suay}
obtained his B.Sc. degree in Mathematics (2007), a Master degree in Advanced Computing and Intelligent Systems (2010) and a Ph.D. degree in Computational Mathematics and Computer Science (2015), all from the Universitat de Val\`encia. He is assistant professor in the Dept. of Mathematics in the Universitat de Val\`encia. He is currently a Postdoctoral Researcher at the Image and Signal Processing (ISP) working on dependence estimation, kernel methods and causal inference for remote sensing data analysis.
\end{IEEEbiography}
% Gus
\begin{IEEEbiography}[{\includegraphics[width=1in,height=1.25in,clip,keepaspectratio]{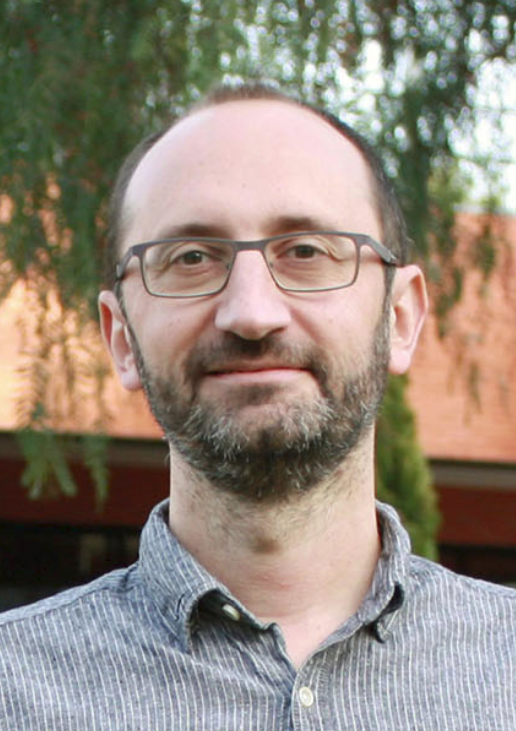}}]{Gustau Camps-Valls}
(M'04, SM'07, FM'18) received a PhD in Physics in 2002 from the Universitat de Val\`encia, and he is currently Full professor in Electrical Engineering, and coordinator of the Image and Signal Processing (ISP) group in the same university, \url{http://isp.uv.es}. He is interested in the development of machine learning algorithms for geoscience and remote sensing data analysis.
\end{IEEEbiography}

% REFERENCES
\bibliographystyle{IEEEtran}
\bibliography{fusion,newbib}

\end{document}